\documentclass[10pt,journal]{IEEEtran}
\usepackage{cite}
\usepackage{graphicx}
\usepackage{subfigure}
\usepackage{algorithm} 
\usepackage{algorithmic}
\usepackage{amsfonts}
\usepackage{amsmath}
\usepackage{bm}
 \usepackage{mathrsfs}
\usepackage{subeqnarray}
\usepackage{cases}
\usepackage{balance}
\usepackage{color}
\usepackage[outdir=./]{epstopdf}

\usepackage{amsmath,lipsum}
\usepackage{cuted}

\begin{document}
%
\title{ISAC from the Sky: UAV Trajectory Design for Joint Communication and Target Localization}
%
%
%

\author{Xiaoye~Jing,~\IEEEmembership{Student Member,~IEEE,}
        Fan Liu,~\IEEEmembership{Member,~IEEE,}
        Christos~Masouros,~\IEEEmembership{Senior Member,~IEEE,}
        Yong Zeng,~\IEEEmembership{Member,~IEEE}\thanks{ 
        This paper has received funding from the European Union’s Horizon 2020 research and innovation programme under the Marie Sklodowska-Curie grant agreement No 812991. {\it{(Corresponding author: Fan Liu.)}}\\
        \indent X. Jing is with the Department of Electronic and Electrical Engineering, University College London, London WC1E 7JE, U.K., and is also with the Department of Electronic and Electrical Engineering, Southern University of Science and Technology, Shenzhen 518055, China (e-mail: x.jing@ucl.ac.uk).\\
        \indent F. Liu is with the Department of Electronic and Electrical Engineering, Southern University of Science and Technology, Shenzhen 518055, China (e-mail: liuf6@sustech.edu.cn).\\
        \indent C. Masouros is with the Department of Electronic and Electrical Engineering, University College London, London WC1E 7JE, U.K. (e-mail: chris.masouros@ieee.org).\\
        \indent Y. Zeng is with the National Mobile Communications Research Laboratory, Southeast University, Nanjing 210096, China, and also with the Purple Mountain Laboratories, Nanjing 211111, China (e-mail: yong\_zeng@seu.edu.cn). }}

\maketitle

\begin{abstract}
Unmanned aerial vehicles (UAVs) as aerial base stations (BSs) are able to provide not only the communication service to ground users, but also the sensing functionality to localize targets of interests. In this paper, we consider an airborne integrated sensing and communications (ISAC) system where a UAV, which acts both as a communication BS and a mono-static radar, flies over a given area to transmit downlink signal to a ground communication user. In the meantime, the same transmitted signal is also exploited for mono-static radar sensing. We aim to optimize the UAV trajectory, such that the performance for both communication and sensing (C$\&$S) is explicitly considered. In particular, we first formulate the trajectory design problem into a weighted optimization problem, where a flexible performance trade-off between C$\&$S is achieved. As a step forward, a multi-stage trajectory design approach is proposed to improve the target estimation accuracy. While the resultant optimization problem is difficult to solve directly, we develop an iterative algorithm to obtain a locally optimal solution. Finally, numerical results show that the target estimation error obtained by the trade-off approach is about an order of magnitude better than a communication-only approach with a slight decrease on communication performance.

\end{abstract}


%
\IEEEpeerreviewmaketitle

\section{Introduction}
\subsection{Background and Motivation}
\IEEEPARstart{N}{ext}-generation wireless networks are expected to go beyond communication services to provide highly accurate and robust sensing services, such as localization, tracking and navigation, to numerous users\cite{b1,b2}. In conventional wireless systems, communication and sensing (C$\&$S) functionalities are usually regarded as separated end-goals, and are thus independently designed, resulting in an inefficient resource utilization. To resolve such issues, designing C$\&$S systems in a joint manner so that they can share the common hardware and signals, is envisioned as a promising solution to reduce the hardware cost and improve spectral efficiency, which has motivated the recent research interest in integrated sensing and communications (ISAC) technology\cite{b1,b2,b3}. ISAC has been widely studied for numerous wireless applications, such as Wi-Fi based indoor localization and cognitive recognition\cite{b4}, cooperative C\&S in unmanned aerial vehicle (UAV) networks\cite{b5} and joint vehicle tracking and communication\cite{b6}.\\
\indent UAV based networks are envisioned as important parts in future wireless systems, which can provide flexible topology and on-demand connectivity in emergency situations\cite{b7,b8,b9}. In particular, UAV based communication systems and UAV based sensing systems are regarded as important components for aerial wireless networks. However, the separate deployment of C$\&$S systems inevitably incurs a heavy payload carried by the UAV, and therefore results in performance loss in the maneuverability of the UAV and a shorter battery lifetime. To that end, ISAC technology is envisioned as a promising solution to minimize the payload of the UAV, enabling the hardware reuse between C\&S\cite{b10}. Moreover, a common waveform can be leveraged for both functionalities, which may significantly increase the spectral and energy efficiency of the UAV network. In the following, we provide a brief overview on the ISAC technology and UAV applications for both C\&S.
\subsection{ISAC Technology}
\indent  Recently, there has been a dramatically growing research interest in ISAC. The initial research effort was to find the appropriate integration framework, from sensing-communication coexistence, which aims at enabling the harmonic cohabitation of the two systems in the same frequency band without interfering with each other\cite{b1}, to dual-functional radar-communication (DFRC) system, which focuses on designing joint systems that can simultaneously perform wireless communication and remote sensing\cite{b1,b11}. In the state-of-the-art DFRC research, sophisticated techniques including coding\cite{b12,b13,b14}, waveform design\cite{b15,b16,b17} and beamforming\cite{b6}, were developed to facilitate the use of a common signal for accomplishing the dual tasks. The performance of the ISAC system can be generally evaluated from both C$\&$S aspects. In particular, for the communication aspect, information and communication theoretic metrics, including channel capacity and symbol error rate\cite{b1,b11}, are typically taken into account, while for the sensing functionalities, estimation theoretic metrics, such as mean-square error (MSE) and Cram\'er-Rao Bound (CRB)\cite{b1,b11} are considered. In some research works, the joint performance metrics for both C\&S could be possibly improved simultaneously. However, in most cases, there are fundamental performance trade-offs between C$\&$S, due to the shared use of wireless and hardware resources, to achieve conflicting design objectives. To that end, it is of particular importance to reveal such performance bounds and to design bound-achieving transmission strategies for ISAC systems. A representative ISAC beamforming design methodology was proposed in \cite{b18}, which minimized the CRB for target sensing while guaranteeing the downlink communications Quality of Service (QoS).  This specific trade-off, exploited for wavefrom design in \cite{b18}, offers the basis of our trajectory design work.
\subsection{UAV Research on C\&S}
Different from the conventional terrestrial cellular networks, where the base stations (BSs) are fixed on the ground, UAV based wireless networks have the additional design degrees of freedom for UAV positioning optimization \cite{b19}. By optimizing the UAV location and its flying status, UAV-based C\&S platform provides the extra degree of freedom to enhance the propagation channel to communication users and sensing targets, resulting in potential performance improvement in C$\&$S services. The research on UAV deployment optimization can be loosely classified into two categories, namely, static UAV deployment and moving UAV trajectory design \cite{b7}.\\
\indent With static UAV deployment for wireless communication, the aim is to find the best UAV hovering positions in space that are optimal for communication. Earlier researches focused on single-UAV communication systems, which aimed to optimize the UAV coverage performance by determining  the UAV's altitude\cite{b20,b21,b22}. To further boost the performance, multi-UAV deployment can be exploited. UAVs' topology and clustering approaches were studied for coverage maximization and energy efficiency improvement \cite{b23,b24,b25,b26,b27,b28,b29}. On the other hand, for sensing, static UAV deployment usually requires a minimum of three UAVs, each can obtain a distance estimation between itself and the target. Based on at least three distance estimations, the target location can be estimated. To simplify the measurement means and exploit the advantage of multi-UAV deployment, static UAV based sensing can be performed over a multi-UAV network\cite{b30}.  \\
\indent Comparing with static UAVs, moving UAVs can fly closer to communication users and sensing targets of interest. Meanwhile, the UAV transmission directions for communication users and targets may evolve with time, improving channel diversity to benefit both C$\&$S functionalities. Since the flying time of the UAV is constrained by its limited battery capacity, energy-efficient UAV communication design has received considerable research attentions. In \cite{b31,b32,b33}, energy efficiency was considered in UAV trajectory design while ensuring the communication performance. In regard to sensing, path design for single-UAV based sensing was studied in \cite{b34}, which considered a 3-dimensional model with closed-form solution for UAV path designs. Angle-of-arrival (AoA) measurement was applied in multi-UAV path design in \cite{b35} and \cite{b36}. In \cite{b36}, a cooperative resource scheduling and a task planning scheme were proposed for both target localization and tracking. \\
\indent  There are some existing researches on UAV based ISAC. The research scenario in \cite{b37} and \cite{b38} includes one UAV, which senses multiple targets and transmits the data to a ground base station for data updating. The objective is to minimize the maximum average of the peak age of information (PAoI) from targets, by jointly optimizing UAV trajectory, task scheduling, time and power allocation. Some UAV based ISAC scenarios focus on the UAV network that is exploited both for ground target sensing and user communication services. In \cite{b5}, a power allocation scheme was proposed in a static multi-UAV network to maximize a utility performance for C$\&$S. The UAV deployment was determined via a clustering approach and the power allocation scheme was studied based on known targets locations. As mentioned above, in order to fully exploit the UAV mobility to improve both C$\&$S performances, a more promising research direction would be ISAC systems with moving UAVs. In \cite{b39}, a periodic ISAC framework was proposed,  where sensing is periodically performed with continuous communication. The authors maximized the communication performance by optimizing the sensing instant, UAV trajectory and beamforming, and at the same time, the trajectory need to satisfy certain sensing requirements. In \cite{b40}, the authors considered a scenario where one UAV sends communication signals to multiple users and senses potential targets at interested areas. Both static UAV and moving UAV scenarios were considered. The objective was to maximize the throughput of communication users by jointly optimizing UAV locations and transmit beamforming, subject to sensing requirements and transmission power constraints. In the aforementioned works on ISAC with moving UAVs, the focus was on improving communication performance, subject to the sensing beampattern gain requirements \cite{b39,b40}. Although the beampattern gain is a valid performance metric for sensing, it fails to reflect the specific target estimation performance, such as the estimation error of the target location. This motivates the study in our work, where instead we focus on the CRB, which directly reflects the target estimation performance \cite{b5,b41}.
\subsection{Contributions}
\indent In this paper, we consider the UAV trajectory design problem for ISAC with moving UAV, which aims to simultaneously improve C$\&$S performances. The objective is to jointly minimize the CRB for target localization and maximize the downlink communication rate for the communication user. Note that existing work on ISAC with moving UAV \cite{b39,b40} only consider the UAV energy consumption related to signal transmission. In practice, however, the UAV energy consumed via flying and hovering is the main energy consumption\cite{b32,b42}. In this work, by considering the limited UAV on-board battery capacity, we impose an energy constraint when designing  the flying trajectory of the UAV.\\
\indent A common transmitted signal is used for both C\&S, where the hardware and signal processing resources are shared between the two functionalities. We assume that the UAV provides downlink communication for a single ground communication user, using the pre-designed ISAC signal following, for example, the approaches of \cite{b11,b12,b13,b14,b15,b16,b18}. In the meantime, UAV performs sensing for a ground target using the same signal. Specifically, the UAV transmits the ISAC signal to the target and receives an echo reflected by the target, based on which, the target location is estimated. The joint C$\&$S performances can be improved by optimizing the UAV trajectory, which is essentially to design UAV flying waypoints and velocities.  \\
\indent We summarize our main contributions as follows:
\begin{itemize}
    \item First, we consider the UAV trajectory design for ISAC systems, where an optimization problem is formulated to simultaneously optimize the downlink communication rate and localization accuracy. We derive the CRB of target's two-dimensional coordinates as the sensing metric. We design the UAV trajectory in ISAC systems by considering the C$\&$S performance trade-off with a weighting factor. By tuning the weighting factor, the UAV trajectory design may flexibly adjust its priority between C\&S.
    \item Second, in reality, the target location is initially unknown to the UAV. To design the accurate UAV trajectory for joint C\&S performance improvement, we propose a multi-stage trajectory design approach to develop an efficient target localization algorithm and verify the superior performance of our proposed trajectory design.
    \item Finally, it is challenging to directly solve the formulated trajectory design problem for the considered ISAC system, due to some non-convex constraints and the non-convex objective function. Successive convex approximation (SCA) technique is then utilized to ﬁnd the convex lower-bound of the non-convex part in the constraints. A descent direction is found for the objective function to obtain a locally optimal solution. An iterative algorithm is proposed for the considered optimization problem.
\end{itemize}
\indent The remainder of this paper is organized as follows. Section \uppercase\expandafter{\romannumeral2} introduces the system model. Section \uppercase\expandafter{\romannumeral3} formulates the UAV trajectory design problem and considers the ISAC protocol for UAV flying process. Section \uppercase\expandafter{\romannumeral4} proposes an algorithm to address the trajectory design problem and a target's coordinates estimation method is developed. In Section \uppercase\expandafter{\romannumeral5}, the proposed approaches are evaluated with numerical results. Finally, the paper is concluded in Section \uppercase\expandafter{\romannumeral6}.

\section{System Model}
In this section, we firstly describe the UAV based ISAC system, where one rotary-wing UAV transmits the ISAC signal during its flying process, to simultaneously provide downlink communication service to the communication user and sense the target. We then elaborate on the performance metrics for C\&S, respectively. 

\subsection{UAV Trajectory Model}
\begin{figure}[t]
\centering
\includegraphics[width=0.95\linewidth]{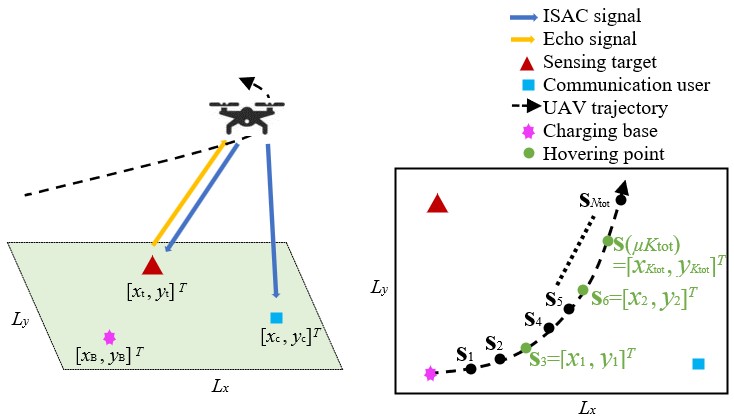}
\caption{Rotary-wing UAV based ISAC system.}
\end{figure}
As shown in Fig. 1, we consider a UAV-based ISAC system, with one rotary-wing UAV, one communication user receiving ISAC signals for its communication service, and one target to be sensed by the UAV. We consider a rectangular area on the ground with a dimension ${L_x}$ by ${L_y}$, and one charging base located at $\left[x_{\rm{B}},y_{\rm{B}}\right]^T\in {\mathbb{R}^{2 \times 1}}$ for UAV to charge its battery. The communication user's location is denoted by $\left[x_{\rm{c}},y_{\rm{c}}\right]^T\in {\mathbb{R}^{2 \times 1}}$. The location information of both the user and the UAV are known {\it{a priori}} via Global Navigation Satellite System. The location of the target is unknown, and is denoted by $\left[x_{\rm{t}},y_{\rm{t}}\right]^T\in {\mathbb{R}^{2 \times 1}}$. \\ 
\indent  The UAV is dispatched from $\left[x_{\rm{B}},y_{\rm{B}}\right]^T$ and then flies with a constant altitude $H$. Its flying path is described by $N_{\rm{tot}}$ waypoints, ${\bf{s}}\left(n\right)\in {\mathbb{R}^{2 \times 1}},n=1,...,N_{\rm{tot}}$, which are concatenated as a matrix ${\bf{S}}=\left[{\bf{s}}\left(1\right),...,{\bf{s}}\left(n\right),...,{\bf{s}}\left(N_{\rm{tot}}\right)\right]\in {\mathbb{R}^{2 \times N_{\rm{tot}}}}$. $N_{\rm{tot}}$ is affected by the UAV on-board battery energy ${E_{{\rm{tot}}}}$. Larger ${E_{{\rm{tot}}}}$ enables the UAV to fly a longer distance with a larger $N_{\rm{tot}}$. The flying path between ${\bf{s}}\left(n-1\right)$ and ${\bf{s}}\left(n\right)$ is a line segment and the average flying velocity in this line segment is 
\begin{equation}
   {\bf{v}}\left(n\right)=\begin{cases}\frac{{\bf{s}}\left(n\right)-{\bf{s}}\left(n-1\right)}{T_{\rm{f}}}, & n=2,...,N_{\rm{tot}},\\
   \frac{{\bf{s}}\left(1\right)-\left[x_{\rm{B}},y_{\rm{B}}\right]^T}{T_{\rm{f}}}, & n=1,
   \end{cases}
\end{equation}
where $T_{\rm{f}}$ is the flying duration  between each pair of consecutive waypoints. We use a matrix ${\bf{V}}=\left[{\bf{v}}\left(1\right),...,{\bf{v}}\left(N_{\rm{tot}}\right)\right]\in {\mathbb{R}^{2 \times N_{\rm{tot}}}}$ to describe the UAV flying velocities along its trajectory.\\
\indent The ISAC signal embedded with useful information is used by the UAV to transmit data continuously to the communication user as it flies following the designed trajectory. Meanwhile, there are $K_{\rm{tot}}$ hovering points included in the $N_{\rm{tot}}$ waypoints for the UAV to perform sensing to the target. Specifically, after the UAV flies over $\mu-1$ waypoints, where $\mu$ is a given integer (in Fig.1, $\mu=3$), it hovers at the next waypoint with a duration $T_{\rm{h}}$ to perform one-time sensing, so that $K_{\rm{tot}}={\rm{floor}}\left(\frac{N_{\rm{tot}}}{\mu}\right)$. The $k$-th hovering point is
\begin{equation}
    \left[x_k,y_k\right]^T={\bf{s}}\left(\mu k\right),k=1,2,...,K_{\rm{tot}}.
\end{equation}
The $K_{\rm{tot}}$ hovering points are denoted by two vectors ${\bf{x}}=\left[x_1,...,x_k,...,x_{K_{\rm{tot}}}\right]^T$ and ${\bf{y}}=\left[y_1,...,y_k,...,y_{K_{\rm{tot}}}\right]^T$. Specifically, at each hovering point, an ISAC signal is transmitted to both the communication user and the target. Then, an echo of the ISAC signal is reflected by the target and received by the UAV. The propagation delay between the ISAC signal and its echo is obtained, based on which the distance between the UAV hovering point and the target is estimated. Finally, these distance measurements from multiple UAV hovering points can then be fused to obtain the horizontal coordinates of the target via estimation method, e.g., maximum likelihood estimation (MLE). We should notice that since the target is on the ground, there may exist reflections from the terrain, which causes interference to the echo. Some filtering  or deconvolution methods for interference suppression can be used here to mitigate the clutter \cite{b43}.

\subsection{Communication Model}
\indent Following the relevant UAV literature \cite{b32}, we assume that the propagation channel is dominated by the light-of-sight (LoS) link, where the channel quality mainly depends on the UAV-receiver distance. Therefore, free-space path loss model is applicable to both C\&S.\\
\indent For the communication model, let $d_{\rm{c}}(n)$ denote the distance from the UAV at the $n$-th waypoint to the communication user, which is
\begin{equation}
    d_{\rm{c}}\left(n\right)=\sqrt{H^2+\|{\bf{s}}\left(n\right)-\left[x_{\rm{c}},y_{\rm{c}}\right]^T\|^2}.
\end{equation}
\indent Therefore, the channel power gain from the UAV at the $n$-th waypoint to the communication user can be expressed as
\begin{equation}
    h\left(n\right)=\frac{\alpha_0}{\left[d_{\rm{c}}\left(n\right)\right]^2},
\end{equation}
where $\alpha_0=\frac{G_{\rm{T}}\cdot G_{\rm{c}} \cdot \lambda^2}{\left(4\pi\right)^2 }$ is the channel power at the reference distance $d_{\rm{c}}\left(n\right)=1{\rm{m}}$. $G_{\rm{T}}$ is the transmit antenna gain. $G_{\rm{c}}$ is the receive antenna gain of the communication user and $\lambda$ is the wavelength. The signal-to-noise ratio (SNR) from the UAV at the $n$-th waypoint to the communication user can be expressed as
\begin{equation}
    SNR_{\rm{c}}(n)=\frac{Ph\left(n\right)}{\sigma^2_0},
\end{equation}
where $P$ is the transmit power. $\sigma^2_0$ is the noise power at the receiver. The user's downlink communication rate at the $n$-th waypoint is
\begin{equation}
    R\left(n\right)=B{\rm{log}}_2\left(1+\frac{Ph\left(n\right)}{\sigma^2_0}\right),
\end{equation}
where $B$ is the channel bandwidth.\\
\indent We use the average downlink communication rate as the communication performance metric. Given the limitations in UAV velocity and with the appropriate choice of duration $T_{\rm{f}}$, the distance between ${\bf{s}}\left(n-1\right)$ and ${\bf{s}}\left(n\right)$ typically satisfies a constraint $\|{\bf{s}}\left(n-1\right)-{\bf{s}}\left(n\right)\| \ll H$. Therefore, the communication distance between the UAV to the communication user within each line segment between two consecutive waypoints ${\bf{s}}\left(n-1\right)$ and ${\bf{s}}\left(n\right)$ is approximately unchanged. Thus, the achievable downlink communication rate when UAV flies from ${\bf{s}}\left(n-1\right)$ to ${\bf{s}}\left(n\right)$ is nearly unchanged. Thus, the average downlink communication rate can be expressed by the average of the communication rate of $N_{\rm{tot}}$ waypoints, which is 
\begin{equation}
     \overline{R}= \dfrac{1}{N_{\rm{tot}}}\sum\limits_{n= 1}^{N_{\rm{tot}}} R\left(n\right).
\end{equation}

\subsection{Radar Sensing Model}
For target location estimation, $\left[x_{\rm{t}},y_{\rm{t}}\right]^T$ can be determined according to at least 3 different distance measurements, which indicates that at least $K_{\rm{tot}}=3$ hovering points are needed. In the sensing model, we denote $\tau\left(k\right)$ as the two-way time delay, propagation delay of the ISAC signal from the UAV at $\left[x_k,y_k\right]^T$ to the target and reflected by the target to the UAV. $d_{\rm{s}}(k)$ is the distance from the UAV to the target, which is
\begin{equation}
    d_{\rm{s}}(k)=\sqrt{H^2+\|\left[x_k,y_k\right]^T-\left[x_{\rm{t}},y_{\rm{t}}\right]^T\|^2}.
\end{equation}
We use a vector ${\bf{d}}_{\rm{s}}=\left[d_{\rm{s}}\left(1\right),...,d_{\rm{s}}\left(K_{\rm{tot}}\right)\right]^T$ to represent distances between the target and all hovering points for convenience. The distance $d_{\rm{s}}\left(k\right)$ is given as
\begin{equation}
   d_{\rm{s}}\left(k\right)=\frac{\tau\left(k\right)\cdot c}{2} ,
\end{equation}
where $c$ is the speed of light. The measurement of $d_{\rm{s}}\left(k\right)$ is 
\begin{equation}
    \widehat{d}_{\rm{s}}\left(k\right)=d_{\rm{s}}\left(k\right)+w_{\tau}\left(k\right),
\end{equation}
where $w_{\tau}\left(k\right)$ denotes the Gaussian noise of the measurement with zero mean and variance of $\sigma^2\left(k\right)$. Note that $\sigma^2\left(k\right)$ is subject to individual Gaussian distribution, which is inversely proportional to the received SNR of the ISAC echo \cite{b44}. We use $\widehat{\bf{d}}_{\rm{s}}=\left[\widehat{d}_{\rm{s}}\left(1\right),...\widehat{d}_{\rm{s}}\left(k\right),...,\widehat{d}_{\rm{s}}\left(K_{\rm{tot}}\right)\right]^T$ for convenience.\\
\indent The two-way channel power gain from the UAV at the $k$-th hovering point $\left[x_k,y_k\right]^T$ to the target and then back to $\left[x_k,y_k\right]^T$ is
\begin{equation}
    g\left(k\right)=\frac{\beta_0}{\left[d_{\rm{s}}\left(k\right)\right]^4},
\end{equation}
where $\beta_0$ is the channel power at the reference distance $d_{\rm{s}}\left(k\right)=1{\rm{m}}$, which is expressed as
\begin{equation}
    \beta_0=\frac{G_{\rm{T}}\cdot G_{\rm{s}}\cdot \sigma_{\rm{rcs}}\cdot \lambda^2}{\left(4\pi\right)^3 },
\end{equation}
where $G_{\rm{s}}$ is the receive antenna gain of the target and $\sigma_{\rm{rcs}}$ is the Radar Cross-Section (RCS). The received SNR of the ISAC echo at $\left[x_k,y_k\right]^T$ is
\begin{equation}
    SNR_{\rm{s}}(k)=\frac{P\cdot G_{\rm{p}}\cdot g\left(k\right)}{\sigma^2_0},
\end{equation}
where $G_{\rm{p}}$ is the signal processing gain.  So that we remark
\begin{equation}
    \sigma^2\left(k\right)= \frac{a\sigma^2_0}{PG_{\rm{p}}g\left(k\right)},
\end{equation}
where $a$ is pre-determined constant related to the system setting. \\
\indent For the target location estimation, the parameters to be estimated are target's two-dimensional coordinates $x_{\rm{t}}$ and $y_{\rm{t}}$, so that ${\bf{u}}=\left[x_{\rm{t}},y_{\rm{t}}\right]^T$.  ${\widehat{\bf{u}}}$ is the estimated vector of $\bf{u}$. To assess the performance of an estimator, the MSE $\epsilon^2=\mathbb{E} \left[\parallel{\bf{u}}-{\widehat{\bf{u}}}\parallel^2\right]$ is a commonly used metric. However, MSE is often difficult to be obtained in closed form and the minimization of MSE is almost intractable. For the unbiased parameter estimator, the CRB can provide a lower bound for MSE\cite{b45}. Therefore, we resort to the CRB instead for sensing performance evaluation. \\
\indent Thus, we need to calculate the CRB of coordinates $x_{\rm{t}}$ and $y_{\rm{t}}$. From \cite{b45}, we know that the CRB of the elements in $\bf{u}$ is included in the CRB matrix of $\bf{u}$, which is denoted by ${\rm{CRB}}_{\bf{u}}$. Specifically, the CRB of the $p$-th element in $\bf{u}$ is the $p$-th diagonal element in ${\rm{CRB}}_{\bf{u}}$, so that the sum of the CRB of coordinates $x_{\rm{t}}$ and $y_{\rm{t}}$ is
\begin{align}
   {\rm{CRB}}_{x_{\rm{t}}}+{\rm{CRB}}_{y_{\rm{t}}}=\left[{\rm{CRB}}_{\bf{u}}\right]_{1,1}+\left[{\rm{CRB}}_{\bf{u}}\right]_{2,2}.   
\end{align}
\indent Note that ${\rm{CRB}}_{\bf{u}}$ can be derived from 
\begin{equation}
     {\rm{CRB}}_{\bf{u}}={\bf{J}}^{-1}\left({\bf{u}}\right),
\end{equation}
where ${\bf{J}}\left({\bf{u}}\right)$ is the Fisher’s information matrix (FIM) of ${\bf{u}}$. \\
\indent In most cases, it is easier to compute  ${\bf{J}}\left({\bf{u}}\right)$  from the FIM based on other related parameters, and then exploit the mathematical relationship between the two FIMs to derive ${\bf{J}}\left({\bf{u}}\right)$. For our considered problem, ${\bf{d}}_{\rm{s}}$ can be estimated based on the echoes received by the UAV, so we can first construct the FIM with respect to ${\bf{d}}_{\rm{s}}$. Then,  ${\bf{J}}\left({\bf{u}}\right)$ can be derived by the chain rule in the form of
\begin{equation}
    {\bf{J}}\left({\bf{u}}\right)={\bf{Q}}{\bf{J}}\left({\bf{d}}_{\rm{s}}\right){\bf{Q}}^T,
\end{equation}
where ${\bf{J}}\left({\bf{d}}_{\rm{s}}\right)$ is the FIM of ${\bf{d}}_{\rm{s}}$ and ${\bf{Q}}$ is a Jacobian matrix in the form of
\begin{equation}
{\bf{Q}}=\frac{\partial{\bf{d}}^T_{\rm{s}}}{\partial{\bf{u}}}=\left[ \begin{array}{cccc}
\frac{x_1-x_{\rm{t}}}{d_{\rm{s}}\left(1\right)}&...&\frac{x_{K_{\rm{tot}}}-x_{\rm{t}}}{d_{\rm{s}}\left({K_{\rm{tot}}}\right)}\\
\frac{y_1-y_{\rm{t}}}{d_{\rm{s}}\left(1\right)}&...&\frac{y_{K_{\rm{tot}}}-y_{\rm{t}}}{d_{\rm{s}}\left({K_{\rm{tot}}}\right)}
\end{array}
\right ].
\end{equation}
\indent We observe that both  $\widehat{d}_{\rm{s}}\left(k\right)$ and  $\sigma^2\left(k\right)$ are dependent on $d_{\rm{s}}\left(k\right)$. Thus, the measurement vector $\widehat{{\bf{d}}}_{\rm{s}}$ follows the distribution of 
\begin{equation}
    \widehat{{\bf{d}}}_{\rm{s}}\sim \mathscr{N} \left({\bf{d}}_{\rm{s}},{\bf{C}}\left({\boldsymbol{d}}_{\rm{s}}\right)\right),
\end{equation}
where ${\bf{C}}\left({\bf{d}}_{\rm{s}}\right)$ is the measurement covariance as
\begin{equation}
    {\bf{C}}\left({\bf{d}}_{\rm{s}}\right)=\frac{a\sigma^2_0}{PG_{\rm{p}}\beta_0}{\rm{diag}}\left(\left[d_{\rm{s}}\left(1\right)\right]^4,\left[d_{\rm{s}}\left(2\right)\right]^4,...,\left[d_{\rm{s}}\left({K_{\rm{tot}}}\right)\right]^4\right).
\end{equation}
Given by \cite{b46} (eq. 3.31), the $p$,$q$-th element in ${\bf{J}}\left({\bf{d}}_{\rm{s}}\right)$ is shown in (21) at the top of the next page.\\
\newcounter{mytempeqncnt}
\begin{figure*}[!t]
\normalsize
\setcounter{mytempeqncnt}{\value{equation}}
\setcounter{equation}{20}
\begin{equation}
    \left[{\bf{J}}\left({\bf{d}}_{\rm{s}}\right)\right]_{pq}=\left[\frac{\partial {\bf{d}}_{\rm{s}}}{\partial d_{\rm{s}}\left(p\right)}\right]^T{\bf{C}}^{-1}\left({\bf{d}}_{\rm{s}}\right)\left[\frac{\partial {\bf{d}}_{\rm{s}}}{\partial d_{\rm{s}}\left(q\right)}\right]+\frac{1}{2}{\rm{tr}}\left[{\bf{C}}^{-1}\left({\bf{d}}_{\rm{s}}\right)\frac{\partial{\bf{C}}\left({\bf{d}}_{\rm{s}}\right)}{\partial d_{\rm{s}}\left(p\right)}{\bf{C}}^{-1}\left({\bf{d}}_{\rm{s}}\right)\frac{\partial{\bf{C}}\left({\bf{d}}_{\rm{s}}\right)}{\partial d_{\rm{s}}\left(q\right)}\right],\ {\rm{where,}}\ p,q=1,...,K_{\rm{tot}}.
\end{equation}
\setcounter{equation}{\value{mytempeqncnt}}
\hrulefill
\vspace*{4pt}
\end{figure*}
\setcounter{equation}{21}
Finally, substituting (18) and (21) into (17), the CRB matrix of ${\bf{u}}$ can be expressed as
\begin{equation}
   {\rm{CRB}}_{\bf{u}}=\left[{\bf{J}}\left({\bf{u}}\right)\right]^{-1}=\frac{1}{\Theta_{\rm{a}}  \Theta_{\rm{b}}-\left[\Theta_{\rm{c}}\right]^2}\left[ \begin{array}{cc}
\Theta_{\rm{b}} & \Theta_{\rm{c}}\\
\Theta_{\rm{c}} & \Theta_{\rm{a}}
\end{array} 
\right ],
\end{equation}
where,
\begin{align}
    \Theta_{\rm{a}}&=\sum_{k=1}^{K_{\rm{tot}}} \left\{\frac{PG_{\rm{p}}\beta_0}{a\sigma^2_0}\times \frac{\left(x_k-x_{\rm{t}}\right)^2}{\left[d_{\rm{s}}\left(k\right)\right]^6}+\frac{8\left(x_k-x_{\rm{t}}\right)^2}{\left[d_{\rm{s}}\left(k\right)\right]^4}\right\},
\end{align}
\begin{align}
    \Theta_{\rm{b}}&=\sum_{k=1}^{K_{\rm{tot}}} \left\{\frac{PG_{\rm{p}}\beta_0}{a\sigma^2_0}\times \frac{\left(y_k-y_{\rm{t}}\right)^2}{\left[d_{\rm{s}}\left(k\right)\right]^6}+\frac{8\left(y_k-y_{\rm{t}}\right)^2}{\left[d_{\rm{s}}\left(k\right)\right]^4}\right\},
\end{align}
\begin{align}
    \Theta_{\rm{c}}=&\sum_{k=1}^{K_{\rm{tot}}} \left\{\frac{PG_{\rm{p}}\beta_0}{a\sigma^2_0}\times \frac{\left(x_k-x_{\rm{t}}\right)\left(y_k-y_{\rm{t}}\right)}{\left[d_{\rm{s}}\left(k\right)\right]^6}\right\}\nonumber\\
    &+\sum_{k=1}^{K_{\rm{tot}}} \left\{\frac{8\left(x_k-x_{\rm{t}}\right)\left(y_k-y_{\rm{t}}\right)}{\left[d_{\rm{s}}\left(k\right)\right]^4}\right\}.
\end{align}\\
\indent The CRB of coordinates $x_{\rm{t}}$ and $y_{\rm{t}}$ are diagonal entries of ${\rm{CRB}}_{\bf{u}}$, which are given by
\begin{equation}
    {\rm{CRB}}_{x_{\rm{t}}}=\frac{ \Theta_{\rm{b}}}{\Theta_{\rm{a}}  \Theta_{\rm{b}}-\left[\Theta_{\rm{c}}\right]^2},
\end{equation}
\begin{equation}
    {\rm{CRB}}_{y_{\rm{t}}}=\frac{ \Theta_{\rm{a}}}{\Theta_{\rm{a}}  \Theta_{\rm{b}}-\left[\Theta_{\rm{c}}\right]^2}.
\end{equation}
We adopt the sum of the CRB of $x_{\rm{t}}$ and $y_{\rm{t}}$ as the sensing performance metric, which is 
\begin{equation}
   {\rm{CRB}}_{x_{\rm{t}},y_{\rm{t}}}= {\rm{CRB}}_{x_{\rm{t}}}+ {\rm{CRB}}_{y_{\rm{t}}}=\frac{ \Theta_{\rm{b}}+\Theta_{\rm{a}}}{\Theta_{\rm{a}}  \Theta_{\rm{b}}-\left[\Theta_{\rm{c}}\right]^2}.
\end{equation}
\indent Finally, the objective of the sensing optimization in the considered ISAC system is to minimize the CRB of $x_{\rm{t}}$ and $y_{\rm{t}}$, which leads to the minimization of the lower bound for the sum of the achievable MSE of $x_{\rm{t}}$ and $y_{\rm{t}}$. 

\section{Problem Formulation and ISAC Protocol}
\subsection{Problem Formulation}
In this section, a UAV trajectory design problem is formulated to jointly optimize the C$\&$S performance.\\
\indent The UAV power consumption consists of the power consumed for the ISAC signal transmission and for supporting its mobility. Nevertheless, the former energy is typically orders of magnitude less than the energy spent for propulsion, and we therefore ignore it in our energy model. Consequently, the propulsion power consumption can be modeled as a function of the velocity as \cite{b32} 
\begin{align}
    P(V)=& P_0\left(1+ \frac{3V^2}{{U_{\rm{tip}}}^2} \right) +P_{\rm{I}}\left(\sqrt{ \left(1+\frac{V^4}{{4v_0}^4}  \right)} - \frac{V^2}{2{v_0}^2} \right)^{\frac{1}{2}}\nonumber\\
    &+\frac{1}{2}D_0\rho s AV^3,
\end{align}
where $P_0$ and $P_{\rm{I}}$ are the blade proﬁle power and induced power in hovering status; ${U_{\rm{tip}}}$ is the tip speed of the rotor blade; $v_0$ is the mean rotor induced velocity in forward flying; $D_0$ is the Fuselage drag ratio; $\rho$ is the air density; $s$ is rotor solidity and $A$ denotes rotor disc area; $V$ is the UAV speed. Typical values of the parameters related with UAV energy consumption model are shown in Table \uppercase\expandafter{\romannumeral1}.\\
\indent We formulate a UAV trajectory design problem to determine the UAV waypoints, hovering points and flying velocities. The objective is to maximize the average communication rate $\bar R$ while minimizing the CRB of target location estimation. We apply a weighting factor $\eta$ to obtain a tractable trade-off between $\overline{R}$ and ${\rm{CRB}}_{x_{\rm{t}},y_{\rm{t}}}$. The optimization problem is formulated as
\begin{subeqnarray}
    &&{\rm{P1}}:\mathop {{\rm{min}}}\limits_{\left\{ {{\bf{S}},\bf{x}},{\bf{y}},{\bf{V}}\right\}} \  \eta\cdot {\rm{CRB}}_{x_{\rm{t}},y_{\rm{t}}}-\left(1-\eta\right)\cdot\overline{R}\nonumber\\
    &&{\rm{s.t.}}\nonumber\\
    &&\|{\bf{v}}\left(n\right)\|\leq V_{\rm{max}},n=1,...,N_{\rm{tot}},\\
    &&0\leq {\bf{S}}\left(1,n\right)\leq L_x,0\leq {\bf{S}}\left(2,n\right)\leq L_y,\\ &&n=1,...,N_{\rm{tot}},\nonumber\\
    && T_{\rm{f}}\cdot\sum\limits_{n = 1}^{N_{\rm{tot}}} P\left(\|{\bf{v}}\left(n\right)\|\right)+T_{\rm{h}}\cdot\sum\limits_{k = 1}^{K_{\rm{tot}}}P\left(0\right) \leq E_{\rm{tot}}.
\end{subeqnarray}
The weighting factor $\eta$ takes values between 0 and 1, and higher $\eta$ values mean that the UAV trajectory design assigns higher priority on the sensing behaviour in the trade-off relationship. $V_{\rm{max}}$ is the maximum UAV available flying speed. Constraint (30b) restricts that the UAV should fly within the given area. Constraint (30c) means that the  UAV energy  consumed over the whole trajectory should be no larger than its total on-board energy.
\begin{table}[t]
\centering
\caption{Energy consumption function parameters}
\label{my-label}
\begin{tabular}{|c|c|c|c|}
\hline 
 \textbf{parameter}&\textbf{value}&\textbf{parameter}&\textbf{value}\\
 \hline 
 $P_0$&80 W&$P_{\rm{I}}$&88.6 W\\
 \hline 
 $U_{\rm{tip}}$&120 m/s& $v_0$&4.03 m/s \\
 \hline
  $D_0$&0.6&$s$&0.05 $\rm{m}^3$\\
 \hline
  $\rho$&1.225 $\rm{kg}/m^3$ &$A$&0.503 $\rm{m}^2$\\
 \hline
\end{tabular}
\end{table}
\subsection{ISAC Protocol}
It is obvious that the sensing performance  ${\rm{CRB}}_{x_{\rm{t}},y_{\rm{t}}}$ depends on the locations of the $K_{\rm{tot}}$ hovering points ${\bf{x}}$, ${\bf{y}}$ and the location of the target $\left[{x}_{\rm{t}},{y}_{\rm{t}}\right]^T$. However, the ground-truth target location  $\left[{x}_{\rm{t}},{y}_{\rm{t}}\right]^T$ is initially unknown. An estimate of $\left[{x}_{\rm{t}},{y}_{\rm{t}}\right]^T$, which is denoted as $\left[\widehat{x}_{\rm{t}},\widehat{y}_{\rm{t}}\right]^T$, is used to substitute the true location $\left[{x}_{\rm{t}},{y}_{\rm{t}}\right]^T$ for evaluating ${\rm{CRB}}_{x_{\rm{t}},y_{\rm{t}}}$ as follows
\begin{align}
    {\rm{CRB}}_{x_{\rm{t}},y_{\rm{t}}}&=\frac{ \Theta_{\rm{b}}+\Theta_{\rm{a}}}{\Theta_{\rm{a}}  \Theta_{\rm{b}}-\left[\Theta_{\rm{c}}\right]^2}\nonumber\\
    &\approx\frac{\widetilde{\Theta}_{\rm{b}}+\widetilde{\Theta}_{\rm{a}}}{\widetilde{\Theta}_{\rm{a}}  \widetilde{\Theta}_{\rm{b}}-\left[\widetilde{\Theta}_{\rm{c}}\right]^2}\nonumber\\
    &=\widetilde{\rm{CRB}} _{x_{\rm{t}},y_{\rm{t}}},
\end{align}
where $\widetilde{\Theta}_{\rm{a}}$, $\widetilde{\Theta}_{\rm{b}}$ and $\widetilde{\Theta}_{\rm{c}}$ are obtained similarly as $\Theta_{\rm{a}}$, $\Theta_{\rm{b}}$ and $\Theta_{\rm{c}}$ in (23)-(25), by replacing  $x_{\rm{t}}$ and $y_{\rm{t}}$ by $\widehat{x}_{\rm{t}}$ and $\widehat{y}_{\rm{t}}$.\\
\indent Note that there may exist a large gap between $ {\rm{CRB}}_{x_{\rm{t}},y_{\rm{t}}}$ and $ \widetilde{\rm{CRB}} _{x_{\rm{t}},y_{\rm{t}}}$, which means that the solution of P1 is based on an inaccurate objective function, leading to  performance loss for trajectory design. In order to successively improve the estimation accuracy, we propose a multi-stage trajectory design approach, which splits the UAV trajectory design problem into several stages. As such, more accurate target estimate and UAV trajectory are gradually obtained.\\
\begin{figure}[t]
\centering
\subfigure[Illustration]{
\begin{minipage}[t]{0.95\linewidth}
\centering
\includegraphics[width=1\linewidth]{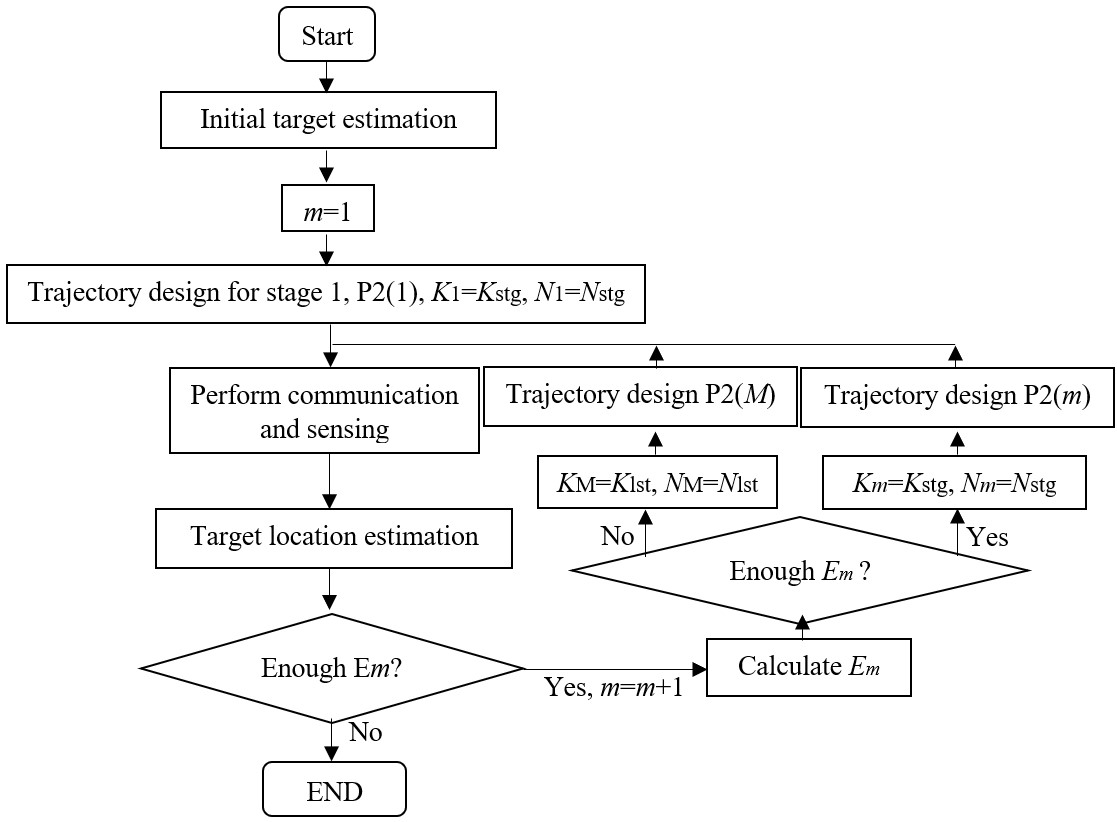}
\end{minipage}
}
\centering
\subfigure[Trajectory]{
\begin{minipage}[t]{0.95\linewidth}
\centering
\includegraphics[width=1\linewidth]{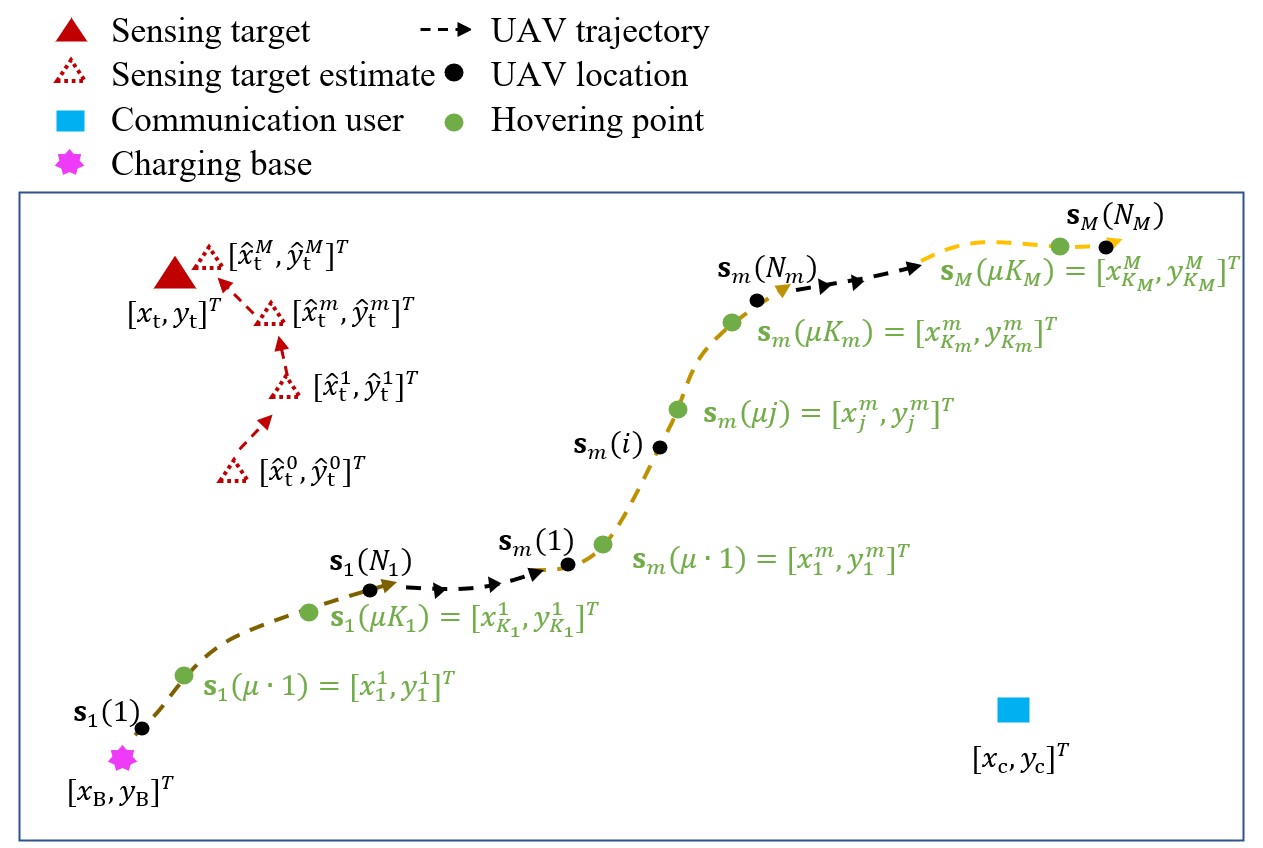}
\end{minipage}
}
\centering
\caption{Multi-stage approach for UAV trajectory design.}
\end{figure}
\indent Denote the number of stages as $M$, which is a variable dependent on the available energy $E_{\rm{tot}}$. In the $m$-th stage ($m=1,2,...,M$), there are $N_m$ UAV waypoints, denoted as ${\bf{S}}_{m}=\left[{\bf{s}}_{m}\left(1\right),...,{\bf{s}}_{m}\left(i\right),...,{\bf{s}}_{m}\left(N_{m}\right)\right]\in \mathbb{R}^{2 \times N_{m}}$. Meanwhile, there are $K_m$ hovering points in the $m$-th stage, denoted as
\begin{equation}
    \left[x^m_j,y^m_j\right]^T={\bf{s}}_m\left(\mu j\right),\,j=1,2,...,K_m.
\end{equation}
We define ${\bf{x}}_m=\left[x^m_1,...,x^m_j,...,x^m_{K_m}\right]^T$ and ${\bf{y}}_m=\left[y^m_1,...,y^m_j,...,y^m_{K_m}\right]^T$ for convenience. After the trajectory of one stage is completed by the UAV, an updated estimate of $\left[{x}_{\rm{t}},{y}_{\rm{t}}\right]^T$, denoted as $\left[\widehat{x}^m_{\rm{t}},\widehat{y}^m_{\rm{t}}\right]^T$, is obtained.\\
\indent Next, we elaborate the details about the multi-stage approach. At the beginning of the approach, we obtain an initial coarse estimate of $\left[{x}_{\rm{t}},{y}_{\rm{t}}\right]^T$, denoted as $\left[\widehat{x}^0_{\rm{t}},\widehat{y}^0_{\rm{t}}\right]^T$. For the first stage $m=1$, we apply $\widehat{x}^0_{\rm{t}}$ and $\widehat{y}^0_{\rm{t}}$ as $\widehat{x}_{\rm{t}}$ and $\widehat{y}_{\rm{t}}$ in $\widetilde{\rm{CRB}} _{x_{\rm{t}},y_{\rm{t}}}$, denoted as $\widetilde{\rm{CRB}}^1 _{x_{\rm{t}},y_{\rm{t}}}$, to formulate the optimization problem for determining the trajectory in the $1$st stage. By solving the trade-off problem, the UAV trajectory and hovering points for the $1$st stage are obtained. In this case, the UAV starts from the initial location $\left[x_{\rm{B}},y_{\rm{B}}\right]^T$ and flies following the obtained path to communicate with the communication user. At each hovering point, the UAV receives an echo from the target. After the UAV completes the $1$th stage, $K_1$ echoes from the sensing target are received, which are used to obtain an updated target location $\left[\widehat{x}^1_{\rm{t}},\widehat{y}^1_{\rm{t}}\right]^T$.\\
\indent Now, the stage index updates to $m=2$. The optimization problem to find the UAV trajectory in the $2$nd stage is formulated. For the objective function of the $m$-th stage, both C$\&$S performance metrics are calculated according to all previous stages plus the new trajectory for the current stage, which is expressed in (33)-(39) at the top of the next page. The CRB is calculated with  $\left[\widehat{x}^{m-1}_{\rm{t}},\widehat{y}^{m-1}_{\rm{t}}\right]^T$, denoted as $\widetilde{\rm{CRB}}^m _{x_{\rm{t}},y_{\rm{t}}}$. After solving the optimization problem for the $m$-th stage,  the UAV starts its flight for the $m$-th stage from the final waypoint of the $\left(m-1\right)$-th stage. \\
\setcounter{equation}{32}
\begin{figure*}[h]
\normalsize
\setcounter{mytempeqncnt}{\value{equation}}
\setcounter{equation}{32}
\begin{align}
    \overline{R}_m= \frac{1}{N_1+N_2+...+N_m}\times \left(\sum_{i=1}^{N_1} B{\rm{log}}_2\left(1+\frac{P\alpha_0}{\sigma^2_0\left[d^1_{\rm{c}}(i)\right]^{2}} \right)+...+\sum_{i=1}^{N_m} B{\rm{log}}_2\left(1+\frac{P\alpha_0}{\sigma^2_0\left[d^m_{\rm{c}}(i)\right]^{2}} \right)\right),
\end{align}
where
\begin{equation}
   {d}^m_{\rm{c}}\left(i\right)=\sqrt{H^2+\|{\bf{s}}_m\left(i\right)-\left[{x}_{\rm{c}},{y}_{\rm{c}}\right]^T\|^2},i=1,2,...,N_m,m=1,2,...,M.
\end{equation} 
\setcounter{equation}{\value{mytempeqncnt}}
\hrulefill
\vspace*{4pt}
\end{figure*}
\setcounter{equation}{34}
\begin{figure*}[h]
\normalsize
\setcounter{mytempeqncnt}{\value{equation}}
\setcounter{equation}{34}
\begin{align}
  \widetilde{\rm{CRB}}^m _{x_{\rm{t}},y_{\rm{t}}}=\frac{\widetilde{\Theta}^m_{\rm{b}}+\widetilde{\Theta}^m_{\rm{a}}}{\widetilde{\Theta}^m_{\rm{a}} \widetilde{\Theta}^m_{\rm{b}}-\left[\widetilde{\Theta}^m_{\rm{c}}\right]^2}.
\end{align}
where
\begin{align}
   \widetilde{\Theta}^m_{\rm{a}}=\sum_{j=1}^{K_1} \left\{\frac{PG_{\rm{p}}\beta_0 } {a\sigma^2_0}\times \frac{\left(x^1_j-\widehat{x}^{m-1}_{\rm{t}}\right)^2}{\left[\widetilde{d}^1_{\rm{s}}\left(j\right)\right]^6}+\frac{8\left(x^1_j-\widehat{x}^{m-1}_{\rm{t}}\right)^2}{\left[\widetilde{d}^1_{\rm{s}}\left(j\right)\right]^4}\right\}+...+\sum_{j=1}^{K_m} \left\{\frac{PG_{\rm{p}}\beta_0 } {a\sigma^2_0}\times \frac{\left(x^m_j-\widehat{x}^{m-1}_{\rm{t}}\right)^2}{\left[\widetilde{d}^m_{\rm{s}}\left(j\right)\right]^6}+\frac{8\left(x^m_j-\widehat{x}^{m-1}_{\rm{t}}\right)^2}{\left[\widetilde{d}^m_{\rm{s}}\left(j\right)\right]^4}\right\}.
\end{align}
\begin{align}
   \widetilde{\Theta}^m_{\rm{b}}=\sum_{j=1}^{K_1} \left\{\frac{PG_{\rm{p}}\beta_0 } {a\sigma^2_0}\times \frac{\left(y^1_j-\widehat{y}^{m-1}_{\rm{t}}\right)^2}{\left[\widetilde{d}^1_{\rm{s}}\left(j\right)\right]^6}+\frac{8\left(y^1_j-\widehat{y}^{m-1}_{\rm{t}}\right)^2}{\left[\widetilde{d}^1_{\rm{s}}\left(j\right)\right]^4}\right\}+...+\sum_{j=1}^{K_m} \left\{\frac{PG_{\rm{p}}\beta_0 } {a\sigma^2_0}\times \frac{\left(y^m_j-\widehat{y}^{m-1}_{\rm{t}}\right)^2}{\left[\widetilde{d}^m_{\rm{s}}\left(j\right)\right]^6}+\frac{8\left(y^m_j-\widehat{y}^{m-1}_{\rm{t}}\right)^2}{\left[\widetilde{d}^m_{\rm{s}}\left(j\right)\right]^4}\right\}.
\end{align}
\begin{align}
   \widetilde{\Theta}^m_{\rm{c}}=&\sum_{j=1}^{K_1} \left\{\frac{PG_{\rm{p}}\beta_0 } {a\sigma^2_0}\times \frac{\left(x^1_j-\widehat{x}^{m-1}_{\rm{t}}\right)\left(y^1_j-\widehat{y}^{m-1}_{\rm{t}}\right)}{\left[\widetilde{d}^1_{\rm{s}}\left(j\right)\right]^6}+\frac{8\left(x^1_j-\widehat{x}^{m-1}_{\rm{t}}\right)\left(y^1_j-\widehat{y}^{m-1}_{\rm{t}}\right)}{\left[\widetilde{d}^1_{\rm{s}}\left(j\right)\right]^4}\right\}+...\nonumber\\
   &+\sum_{j=1}^{K_m} \left\{\frac{PG_{\rm{p}}\beta_0 } {a\sigma^2_0}\times \frac{\left(x^m_j-\widehat{x}^{m-1}_{\rm{t}}\right)\left(y^m_j-\widehat{y}^{m-1}_{\rm{t}}\right)}{\left[\widetilde{d}^m_{\rm{s}}\left(j\right)\right]^6}+\frac{8\left(x^m_j-\widehat{x}^{m-1}_{\rm{t}}\right)\left(y^m_j-\widehat{y}^{m-1}_{\rm{t}}\right)}{\left[\widetilde{d}^m_{\rm{s}}\left(j\right)\right]^4}\right\},
\end{align}
\begin{equation}
   \widetilde{d}^m_{\rm{s}}\left(j\right)=\sqrt{H^2+\left(x^m_j-\widehat{x}^{m-1}_{\rm{t}}\right)^2+\left(y^m_j-\widehat{y}^{m-1}_{\rm{t}}\right)^2},j=1,2,...,K_m,m=1,2,...,M.
\end{equation} 
\setcounter{equation}{\value{mytempeqncnt}}
\hrulefill
\vspace*{4pt}
\end{figure*}
\setcounter{equation}{39}
We assume that, except the final stage, $N_m$ is the same for all stages,  so that $K_m$ is the same as well, which are pre-determined as $N_m=N_{\rm{stg}}$ and $K_m=K_{\rm{stg}}={\rm{floor}}\left(\frac{N_{\rm{stg}}}{\mu}\right),m=1,2,...M-1$. We notice that, in the final stage, the UAV's remaining energy may not be enough to support the UAV to complete a flight with $N_{\rm{stg}}$ waypoints. Therefore, we calculate the maximum achievable $N_M$ in the final stage, and denote it by $N_M=N_{\rm{lst}}$. Meanwhile, there are $K_M=K_{\rm{lst}}={\rm{floor}}\left(\frac{N_{\rm{lst}}}{\mu}\right)$ hovering points. ${\bf{V}}_m=\left[{\bf{v}}_m\left(1\right),...,{\bf{v}}_m\left(i\right),...,{\bf{v}}_m\left(N_m\right)\right]\in{\mathbb{R}^{2 \times N_m}}$ are the UAV velocities in the $m$-th stage. $E_m$ is the UAV's remaining energy for the $m$-th stage ( $E_m=E_{\rm{tot}}$ in the $1$st stage). The multi-stage process is repeated until energy runs out. Finally, we set $K_1+K_2+...+K_M=K_{\rm{tot}}$ and $N_1+N_2+...+N_M=N_{\rm{tot}}$ for later use.\\
\indent Consequently, the trajectory design problem for the $m$-th stage is reformulated as
\begin{subeqnarray}
    &&\mathop{\rm{P2}}\left(m\right):\mathop {{\rm{min}}}\limits_{\left\{ {\bf{S}}_m,{\bf{x}}_m,{\bf{y}}_m,{\bf{V}}_m\right\}} \eta\cdot\widetilde{\rm{CRB}}^m _{x_{\rm{t}},y_{\rm{t}}}-\left(1-\eta\right)\cdot \overline{R}_m\nonumber\\
    &&{\rm{s.t.}}\nonumber\\
    &&\|{\bf{v}}_m\left(i\right)\|\leq V_{\rm{max}},i=1,...,N_m,\\
    &&0\leq {\bf{S}}_m\left(1,i\right)\leq L_x,0\leq {\bf{S}}_m\left(2,i\right)\leq L_y,\\ &&i=1,...,N_m,\nonumber\\
    && T_{\rm{f}}\cdot\sum\limits_{i = 1}^{N_m} P\left(\|{\bf{v}}_m\left(i\right)\|\right)+T_{\rm{h}}\cdot\sum\limits_{j = 1}^{K_m}P\left(0\right) \leq E_m.
\end{subeqnarray}

\section{Proposed algorithm for solving ${\rm{P2}}\left(m\right)$}
\indent In Section \uppercase\expandafter{\romannumeral3}, we formulated the UAV flying process as a multi-stage problem, where the trajectory design and target estimation are proceed in a successive manner. In each stage, the UAV trajectory design is formulated as ${\rm{P2}}\left(m\right)$. However, ${\rm{P2}}\left(m\right)$ is difficult to solve due to the non-convex objective and constraints. In this section, an iterative algorithm is proposed  for solving ${\rm{P2}}\left(m\right)$. Then, an estimation method based on MLE is developed for practical target's coordinates estimation.
\subsection{Proposed Iterative Algorithm}
\indent Due to the non-convex objective function and constraints in ${\rm{P2}}\left(m\right)$, it is difficult to obtain a globally optimal solution. Here we propose an iterative algorithm to address ${\rm{P2}}\left(m\right)$ and seek for its locally optimal solution.\\
\indent We first deal with the non-convex constraint (40c). Since the second term of the left-hand-side (LHS) of (29) is non-convex, we firstly introduce variables ${\bm{\delta}}_m=\{\delta_m\left(1\right),...,\delta_m\left(N_m\right)\}$ to recast this term as
\begin{align}
  \left[\delta_m\left(i\right)\right]^2=\sqrt{ \left(1+\frac{\|{\bf{v}}_m\left(i\right)\|^4}{4v_0^4}  \right)} - \frac{\|{\bf{v}}_m\left(i\right)\|^2}{2v_0^2},i=1,...,N_m,
\end{align}
\begin{equation}
    \delta_m\left(i\right)\geq 0,i=1,...,N_m,
\end{equation}
and the equation (41) above is equivalent to 
\begin{equation}
    \frac{\|{\bf{v}}_m\left(i\right)\|^2}{v_0^2}=\frac{1}{\left[\delta_m\left(i\right)\right]^2}-\left[\delta_m\left(i\right)\right]^2,i=1,...,N_m.
\end{equation}
Based on (41), the constraint (40c) can be rewritten as
\begin{align}
E_m&\geq T_{\rm{f}}\sum\limits_{i = 1}^{N_m} \left\{P_0\left(1+ \frac{3\|{\bf{v}}_m\left(i\right)\|^2}{{U_{\rm{tip}}}^2} \right)+\frac{1}{2}D_0\rho s A\|{\bf{v}}_m\left(i\right)\|^3\right\}\nonumber\\
&+T_{\rm{f}}\sum\limits_{i = 1}^{N_m} P_{\rm{I}} \delta_m\left(i\right)+T_{\rm{h}}\sum\limits_{j = 1}^{K_m} \left\{P_0+P_{\rm{I}}\right\},
\end{align}
with new constraints (42) and (43).  With the above manipulations, ${\rm{P2}}\left(m\right)$ can be written as
\begin{align}
    &{\rm{P2'}}\left(m\right):\mathop {{\rm{min}}}\limits_{\left\{ {\bf{S}}_m, {\bf{x}}_m,{\bf{y}}_m,{\bf{V}}_m\right\}}\eta\cdot\widetilde{\rm{CRB}}^m _{x_{\rm{t}},y_{\rm{t}}}-\left(1-\eta\right)\cdot\overline{R}_m \nonumber\\
    &{\rm{s.t.}}\nonumber\\
    &\frac{\|{\bf{v}}_m\left(i\right)\|^2}{v_0^2}\geq\frac{1}{\left[ \delta_m\left(i\right)\right]^2}-\left[ \delta_m\left(i\right)\right]^2, i=1,...,N_m.\\
    &{\rm{(40a),\ (40b),\ (42)\ and\ (44)}}.\nonumber
\end{align}
Note that (45) is obtained by replacing the equality constraints (43) with inequality constraint, which does not affect the equivalence between ${\rm{P2}}\left(m\right)$ and ${\rm{P2'}}\left(m\right)$. Specifically, suppose that if (45) is satisfied with strict inequality, then we can reduce the value of $\delta_m\left(i\right)$ to enforce that (45) is satisfied with equality. Meanwhile, this reduces the energy consumption function and results in a decrease of the objective value. Consequently, all constraints in (45) will be satisfied with equality, which ensures the equivalence between ${\rm{P2}}\left(m\right)$ and ${\rm{P2'}}\left(m\right)$.\\
\indent ${\rm{P2'}}\left(m\right)$ is still non-convex due to (45). We introduce variables ${\bm{\xi}}_m=\{\xi_m\left(1\right),...,\xi_m\left(i\right),...,\xi_m\left(N_m\right)\}$ and rewrite (45) as
\begin{equation}
    \frac{\|{\bf{v}}_m\left(i\right)\|^2}{v_0^2}\geq\frac{1}{\left[\delta_m\left(i\right)\right]^2}-\xi_m\left(i\right),i=1,2,...,N_m,
\end{equation}
\begin{equation}
      \left[\delta_m\left(i\right)\right]^2\geq\xi_m\left(i\right),i=1,2,...,N_m,
\end{equation}
\begin{equation}
      \xi_m\left(i\right)\geq 0,i=1,2,...,N_m.
\end{equation}
We observe that at the optimal solution to ${\rm{P2'}}\left(m\right)$, we must have $\left[\delta_m\left(i\right)\right]^2= \xi_m\left(i\right)$. We note the inequality constraint in (47), which by a process similar to the above (reduce the value of $\delta_m\left(i\right)$), can be satisfied with equality. Therefore, ${\rm{P2'}}\left(m\right)$ with constraints (46)-(48) is equivalent to ${\rm{P2'}}\left(m\right)$ with constraint (45).\\
\indent Problem ${\rm{P2'}}\left(m\right)$ is still non-convex due to the newly introduced non-convex constraints (46) and (47). However, constraints (46) and (47) can be addressed with SCA algorithm. Specifically, since any convex function is globally lower-bounded by its first-order Taylor expansion at any point \cite{b47}, we first approximate the LHS of (46) by its first-order Taylor expansion near a given point ${\bf{v}}_m^{(l-1)}\left(i\right)$ from the last iteration as
\begin{align}
      &\frac{\|{\bf{v}}_m\left(i\right)\|^2}{v_0^2}\nonumber\\
      &\geq\frac{\|{\bf{v}}_m^{(l-1)}\left(i\right)\|^2}{v_0^2}+ \frac{2}{v_0^2}\left[{\bf{v}}_m^{(l-1)}\left(i\right)\right]^T\left({\bf{v}}_m\left(i\right)-{\bf{v}}_m^{(l-1)}\left(i\right)\right)\nonumber\\
      &\geq\frac{1}{\left[\delta_m\left(i\right)\right]^2}-\xi_m\left(i\right).
\end{align}
Thus, we find the lower bound for the LHS of (46). Then, we approximate the LHS of (47) by its first-order Taylor expansion near a given point $\delta_m^{(l-1)}\left(i\right)$ as
\begin{align}
     &\left[\delta_m\left(i\right)\right]^2\nonumber\\
     &\geq \left[\delta_m^{(l-1)}\left(i\right)\right]^2+2\delta_m^{(l-1)}\left(i\right)\left(\delta_m\left(i\right)-\delta_m^{(l-1)}\left(i\right)\right)\nonumber\\
     &\geq\xi_m\left(i\right).
\end{align}
So that the lower bound of the LHS of (47) is revealed. \\
\indent Now, we reformulate ${\rm{P2'}}\left(m\right)$ as
\begin{subeqnarray}
    &&{\rm{P2''}}\left(m\right):\mathop {{\rm{min}}}\limits_{\left\{ {\bf{S}}_m, {\bf{x}}_m,{\bf{y}}_m,{\bf{V}}_m\right\}}\eta\cdot\widetilde{\rm{CRB}}^m _{x_{\rm{t}},y_{\rm{t}}}-\left(1-\eta\right)\cdot\overline{R}_m \nonumber\\
    &&{\rm{s.t.}}\nonumber\\
    &&\frac{\|{\bf{v}}_m^{(l-1)}\left(i\right)\|^2}{v_0^2}+ \frac{2}{v_0^2}\left[{\bf{v}}_m^{(l-1)}\left(i\right)\right]^T\left({\bf{v}}_m\left(i\right)-{\bf{v}}_m^{(l-1)}\left(i\right)\right)\nonumber\\
    &&\geq\frac{1}{\left[\delta_m\left(i\right)\right]^2}-\xi_m\left(i\right),i=1,...,N_m,\\
    &&\left[\delta_m^{(l-1)}\left(i\right)\right]^2+2\delta_m^{(l-1)}\left(i\right)\left(\delta_m\left(i\right)-\delta_m^{(l-1)}\left(i\right)\geq\xi_m\left(i\right)\right),\nonumber\\
    &&i=1,...,N_m,\\
    &&{\rm{(40a),\ (40b),\ (42),\ (44)\ and\ (48)}}.\nonumber
\end{subeqnarray}
Note that due to the global lower bounds in (51a) and (51b), if the constraints of problem ${\rm{P2''}}\left(m\right)$ are satisfied, then those for the original problem ${\rm{P2}}\left(m\right)$ are guaranteed to
be satisfied as well.\\ 
\indent Finally, all constraints in ${\rm{P2''}}\left(m\right)$ are convex now. But ${\rm{P2''}}\left(m\right)$ is still non-convex because of the non-convex objective term $\widetilde{\rm{CRB}}^m _{x_{\rm{t}},y_{\rm{t}}}$. Now, we illustrate the iterative algorithm to tackle the non-convexity. We denote $\widetilde{\rm{CRB}}^m _{x_{\rm{t}},y_{\rm{t}}}$ as
\begin{equation}
    \Phi\left({\bf{x}}_1,...,{\bf{x}}_m,{\bf{y}}_1,...,{\bf{y}}_m\right)=\widetilde{\rm{CRB}}^m _{x_{\rm{t}},y_{\rm{t}}}.
\end{equation}
At first, we approximate $ \Phi\left({\bf{x}}_1,...,{\bf{x}}_m,{\bf{y}}_1,...,{\bf{y}}_m\right)$  by its first-order Taylor expansion near the given points ${\bf{x}}_1^{(l-1)},...,{\bf{x}}_m^{(l-1)}$ and ${\bf{y}}_1^{(l-1)},...,{\bf{y}}_m^{(l-1)}$  from the last SCA iteration as (53) at the top of next page, where $\nabla \Phi_{x^m_j}$ and $\nabla \Phi_{y^m_j}$ represent the gradient of $\Phi\left({\bf{x}}_1,...,{\bf{x}}_m,{\bf{y}}_1,...,{\bf{y}}_m\right)$ with respect to $x^m_j$ and $y^m_j$ respectively. We denote the last two terns of the right-hand-side (RHS) of (53) as $ F\left({\bf{x}}_1,...,{\bf{x}}_m,{\bf{y}}_1,...,{\bf{y}}_m\right)$.\\
\begin{figure*}[t]
\normalsize
\setcounter{mytempeqncnt}{\value{equation}}
\setcounter{equation}{52}
\begin{align}
  \Phi\left({\bf{x}}_1,...,{\bf{x}}_m,{\bf{y}}_1,...,{\bf{y}}_m\right)\approx&\Phi\left({\bf{x}}_1^{(l-1)},...,{\bf{x}}_m^{(l-1)},{\bf{y}}_1^{(l-1)},...,{\bf{y}}_m^{(l-1)}\right)\nonumber\\
  &+\sum\limits_{j = 1}^{K_m}\nabla \Phi_{x_j^m}\left({\bf{x}}_1^{(l-1)},...,{\bf{x}}_m^{(l-1)},{\bf{y}}_1^{(l-1)},...,{\bf{y}}_m^{(l-1)}\right)\left(x_j^m-\left[x_j^m\right]^{(l-1)}\right)\nonumber\\
  &+\sum\limits_{j = 1}^{K_m}\nabla \Phi_{y_j^m}\left({\bf{x}}_1^{(l-1)},...,{\bf{x}}_m^{(l-1)},{\bf{y}}_1^{(l-1)},...,{\bf{y}}_m^{(l-1)}\right)\left(y_j^m-\left[y_j^m\right]^{(l-1)}\right),
\end{align}
\setcounter{equation}{\value{mytempeqncnt}}
\hrulefill
\vspace*{4pt}
\end{figure*}
\setcounter{equation}{53}
At the $l$-th iteration of the algorithm, we rewrite ${\rm{P2''}}\left(m\right)$ as
\begin{align}
    &{\rm{P2'''}}\left(m\right):\nonumber\\
    &\mathop {{\rm{min}}}_{\left\{{\bf{S}}_m, {\bf{x}}_m,{\bf{y}}_m,
    \atop
    {\bf{V}}_m,{\bm{\delta}}_m,{\bm{\xi}}_m\right\}} \  \eta F\left({\bf{x}}_1,...,{\bf{x}}_m,{\bf{y}}_1,...,{\bf{y}}_m\right)-\left(1-\eta\right)\overline{R}_m\nonumber\\
    &{\rm{s.t.\ }}{\rm{\ (40a),\ (40b),\ (42),\ (44),\ (48),\ (51a)\ and\ (51b)}}.\nonumber
\end{align}
\indent The optimal solution of $\bf{x}_m$ and $\bf{y}_m$ in ${\rm{P2'''}}\left(m\right)$ are denoted as $\bf{x}_m^{\ast}$ and $\bf{y}_m^{\ast}$ respectively. Since the optimization problem is a minimization problem, in the $l$-th  iteration, ${\bf{x}}_m^{\ast}-{\bf{x}}_m^{(l-1)}$ and ${\bf{y}}_m^{\ast}-{\bf{y}}_m^{(l-1)}$ are descent directions of $\bf{x}_m$ and $\bf{y}_m$ \cite{b47}. Next, we set different values of stepsize $\omega_m\ (0\leq \omega_m \leq 1)$ and use them in order to seek for points along ${\bf{x}}_m^{\ast}-{\bf{x}}_m^{(l-1)}$ and ${\bf{y}}_m^{\ast}-{\bf{y}}_m^{(l-1)}$ to find the optimal $\omega_m^{\ast}$ that can obtain the minimum $ \Phi\left({\bf{x}}_1,...,{\bf{x}}_m,{\bf{y}}_1,...,{\bf{y}}_m\right)$. The result of $\bf{x}_m$ and $\bf{y}_m$ are expressed respectively as
\begin{equation}
    {\bf{x}}_m={\bf{x}}_m^{(l-1)}+\omega_m^{\ast}\left({\bf{x}}_m^{\ast}-{\bf{x}}_m^{(l-1)}\right),
\end{equation}
\begin{equation}
    {\bf{y}}_m={\bf{y}}_m^{(l-1)}+\omega_m^{\ast}\left({\bf{y}}_m^{\ast}-{\bf{y}}_m^{(l-1)}\right).
\end{equation}
To guarantee that ${\bf{x}}_m$ and ${\bf{y}}_m$ obtained from every iteration satisfy all constraints in ${\rm{P2'''}}\left(m\right)$, we need to prove that all possible solutions of $\left\{{\bf{S}}_m, {\bf{x}}_m,{\bf{y}}_m,{\bf{V}}_m,{\bm{\delta}}_m,{\bm{\xi}}_m\right\}$ belonging to the descent directions are in the feasible region of the problem. The relevant proof is illustrated in Appendix A.
\subsection{Estimation Approach for Target Location}
\begin{figure*}[t]
\normalsize
\setcounter{mytempeqncnt}{\value{equation}}
\setcounter{equation}{55}
\begin{equation}
   \widehat{\bf{D}}_m=\left[\widehat{d}^1_{\rm{s}}\left(1\right),...,\widehat{d}^1_{\rm{s}}\left(K_1\right),...,\widehat{d}^m_{\rm{s}}\left(1\right),...,\widehat{d}^m_{\rm{s}}\left(K_m\right)\right]^T, {\rm{where\ }} \widehat{d}^m_{\rm{s}}\left(j\right) {\rm{is\ the\ measurement\ of\ }}d^m_{\rm{s}}\left(j\right).
\end{equation}
\setcounter{equation}{\value{mytempeqncnt}}
\hrulefill
\vspace*{4pt}
\end{figure*}
\setcounter{equation}{56}
\begin{figure*}[t]
\normalsize
\setcounter{mytempeqncnt}{\value{equation}}
\setcounter{equation}{56}
\begin{align}
    p\left(\widehat{\bf{D}}_m;\left[x_{\rm{t}},y_{\rm{t}}\right]\right)=&\prod_m\frac{1}{\sqrt{2\pi\left[\sigma_m\left(j\right)\right]^2}}{\rm{exp}}\left[-\frac{1}{2\left[\sigma_m\left(j\right)\right]^2}\left(\widehat{d}^m_{\rm{s}}\left(j\right)-\sqrt{\left(x^m_j-x_{\rm{t}}\right)^2+\left(y^m_j-y_{\rm{t}}\right)^2+H^2}\right)^2\right],
\end{align}
\setcounter{equation}{\value{mytempeqncnt}}
\hrulefill
\vspace*{4pt}
\end{figure*}
\setcounter{equation}{57}
\indent Now, we illustrate the estimation approach for the target's coordinate estimation. The estimate of the target's coordinates $\left[x_{\rm{t}},y_{\rm{t}}\right]^T$ can be obtained via MLE \cite{b45}. In the simulation results that follows, we illustrate the performance of the MLE with respect to CRB.\\
\indent As discussed in Section \uppercase\expandafter{\romannumeral3}, the UAV trajectory design problem is a multi-stage process. At the end of the $m$-th stage, the measurement vector contains $K_1+...+K_m$ elements, which is expressed as (56) at the top of the next page.\\
\indent The likelihood function of $\left[x_{\rm{t}},y_{\rm{t}}\right]^T$ with respect to the measurement vector is shown in (57) at the top of the next page. The MLE of $\left[x_{\rm{t}},y_{\rm{t}}\right]^T$ is defined to be the value that maximizes the likelihood function, which is expressed as
\begin{equation}
    \left[\widehat{x}^m_{\rm{t}},\widehat{y}^m_{\rm{t}}\right]= {\rm{arg}}\left\{\underset{\left[x_{\rm{t}},y_{\rm{t}}\right]}{\rm{max}}\left[{\rm{log}}\ p\left(\widehat{\bf{D}}_m;\left[x_{\rm{t}},y_{\rm{t}}\right]\right)\right]\right\}.
\end{equation}
Thus, the MLE is obtained by letting the derivative equal to zero, i.e.,
\begin{equation}
    \left.\frac{\partial}{\partial \left[x_{\rm{t}},y_{\rm{t}}\right]}{\rm{log}}\ p\left(\widehat{\bf{D}}_m;\left[x_{\rm{t}},y_{\rm{t}}\right]\right)\right|_{\left[x_{\rm{t}},y_{\rm{t}}\right]=\left[\widehat{x}^m_{\rm{t}},\widehat{y}^m_{\rm{t}}\right]}=0.
\end{equation}
with the second-order derivative is less than 0. Since a closed-form expression is unobtainable for the MLE in (59), numerical methods need to be employed. A grid search is applied here in $p\left(\widehat{\bf{D}}_m;\left[x_{\rm{t}},y_{\rm{t}}\right]\right)$ to find $\left[\widehat{x}^m_{\rm{t}},\widehat{y}^m_{\rm{t}}\right]^T$ that can maximize $p\left(\widehat{\bf{D}}_m;\left[x_{\rm{t}},y_{\rm{t}}\right]\right)$. Finally, the MSE of the MLE is evaluated using the Monte Carlo simulation. The proposed approach for UAV trajectory design is described in Algorithm 1. $E_{\rm{min}}$ is the energy consumed for a path with $N_{\rm{stg}}$ waypoints.
\begin{algorithm}[t!]
\caption{The proposed multi-stage approach for UAV trajectory design}
\begin{algorithmic}[1]
\renewcommand{\algorithmicrequire}{\textbf{Initialization:}} 
\REQUIRE
initial target estimation; ${\rm{P2\left(1\right)}}$ for stage 1; $m = 1$; $E_m= E_{\rm{tot}}$
\REPEAT
\STATE
$K_m=K_{\rm{stg}}$ and $N_m=N_{\rm{stg}}$
\STATE
{\rm{for}}\ $j=1:K_m$
\STATE
\,\,\,\,\,\,\,\,\,\,UAV hovers at $\left[x^m_j,y^m_j\right]^T$ to
transmit signal,
\STATE
\,\,\,\,\,\,\,\,\,\,UAV obtains $\widehat{d}^m_{\rm{s}}\left(j\right)$, 
\STATE
end
\STATE
target location estimation $\left[\widehat{x}^m_{\rm{t}},\widehat{y}^m_{\rm{t}}\right]^T$ via MLE,
\STATE
update $m = m + 1$,
\STATE
calculate $E_m$,
\STATE
UAV trajectory design via ${\rm{P2}}\left(m\right)$,
\UNTIL $E_m\leq E_{\rm{min}}$.
\STATE
calculate $N_{\rm{lst}}$ and $K_{\rm{lst}}$ depending on $E_m$,
\STATE
$K_M=K_{\rm{lst}}$ and $N_M=N_{\rm{lst}}$,
\STATE
UAV trajectory design via ${\rm{P2}}\left(M\right)$,
\STATE
{\rm{for}}\ $j=1:K_{\rm{lst}}$
\STATE
\,\,\,\,\,\,\,\,\,\,UAV hovers at $\left[x^M_j,y^M_j\right]^T$ to
transmit signal,
\STATE
\,\,\,\,\,\,\,\,\,\,UAV obtains $\widehat{d}^M_{\rm{s}}\left(j\right)$.
\STATE
end
\STATE
target location estimation $\left[\widehat{x}^M_{\rm{t}},\widehat{y}^M_{\rm{t}}\right]^T$ via MLE.
\end{algorithmic}
\end{algorithm}
\begin{figure}[t!]
\centering
\includegraphics[width=0.8\linewidth]{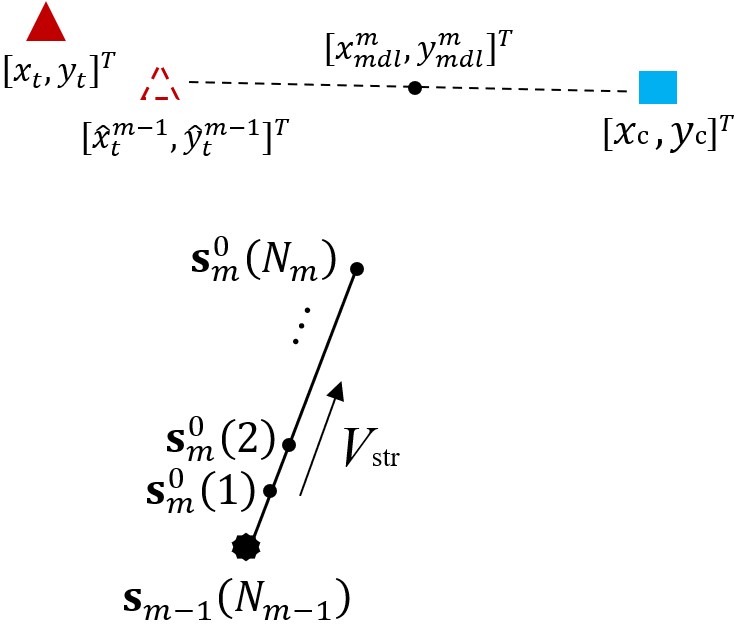}
\caption{Designed trajectory for initial inputs.}
\end{figure}
\subsection{Initialization of the Iterative Algorithm}
The iterative algorithm applied in ${\rm{P2'''}}\left(m\right)$ is an iterative process, thus we need the initial inputs ${\bf{S}}_m^0$, ${\bf{x}}_m^0$, ${\bf{y}}_m^0$, ${\bf{V}}_m^0$, ${\bm{\delta}}_m^0$, and ${\bm{\xi}}_m^0$ for iteration index $l=1$ to start the iterative process. To determine a reasonable set of initial inputs, we design a naive initial trajectory as follows, based on which, ${\bf{S}}_m^0$, ${\bf{x}}_m^0$, ${\bf{y}}_m^0$, ${\bf{V}}_m^0$, ${\bm{\delta}}_m^0$, and ${\bm{\xi}}_m^0$ are calculated. For the $m$-th stage, we use a straight line to connect  $\left[\widehat{x}^{m-1}_{\rm{t}},\widehat{y}^{m-1}_{\rm{t}}\right]^T$ with the communication user and set the middle point of this line as $\left[x^m_{\rm{mdl}},y^m_{\rm{mld}}\right]^T$. To pursue a relative fairness between the C$\&$S functions, the trajectory shown in Fig. 3 is a straight line and its direction is from the start point to $\left[x^m_{\rm{mdl}},y^m_{\rm{mld}}\right]^T$ with a fixed flying speed $V_{\rm{str}}$.  Thus, in the $m$-th stage, the designed trajectory for initial inputs calculation is designed as
\begin{align}
    {\bf{s}}^0_m\left(i\right)&={\bf{s}}_{m-1}\left(N_{m-1}\right)\nonumber\\
    &+\frac{V_{\rm{str}} \cdot i\cdot \left(\left[x^m_{\rm{mdl}},y^m_{\rm{mld}}\right]^T-{\bf{s}}_{m-1}\left(N_{m-1}\right)\right)}{\left\|\left[x^m_{\rm{mdl}},y^m_{\rm{mld}}\right]^T-{\bf{s}}_{m-1}\left(N_{m-1}\right)\right\|}.
\end{align}
where ${\bf{s}}^0_m\left(i\right)$ is the initial input of the $i$-th UAV waypoint in the $m$-th stage for the algorithm. ${\bf{s}}_{m-1}\left(N_{m-1}\right)$ is the last UAV waypoint in the $\left(m-1\right)$-th stage. As such, the initial inputs can be calculated.

\section{simulation results}
\begin{figure}[t]
\centering
\includegraphics[width=0.7\linewidth]{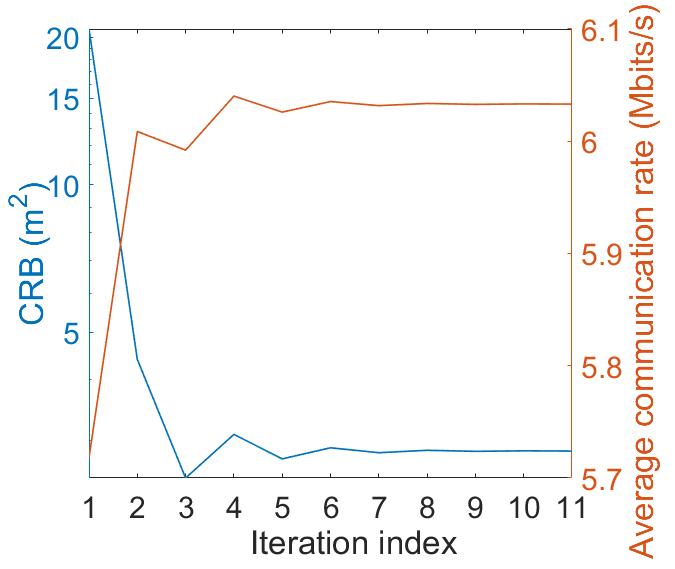}
\caption{Convergence behaviour of the proposed algorithm in the $1$st stage ($N_{\rm{stg}}=25$, $\eta=0.8$ and $E_{{\rm{tot}}}=35{\rm{KJ}}$).}
\end{figure}
\begin{figure}[t]
\centering
\includegraphics[width=0.7\linewidth]{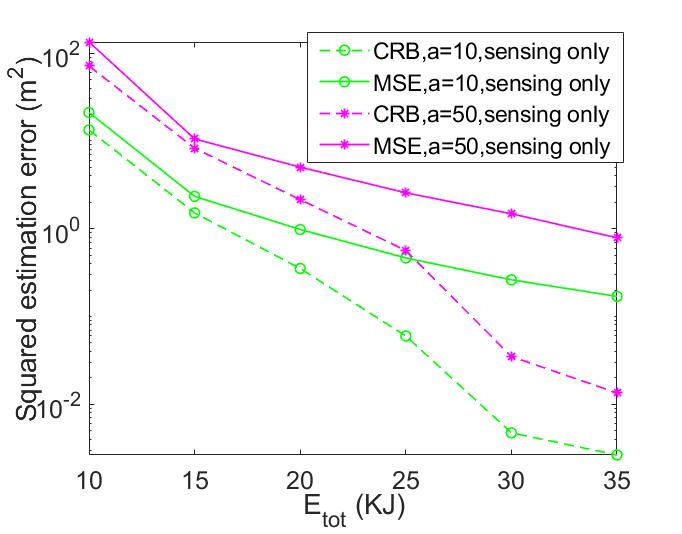}
\caption{Sensing performance of the proposed trajectory design approach based on different $a$ ($N_{\rm{stg}}=25$).}
\end{figure}
This section provides numerical results to evaluate the performance of our proposed approaches. We set the location of the charging base for the UAV to charge its battery as  $\left[x_{\rm{B}},y_{\rm{B}}\right]^T=\left[100{\rm{m}},100{\rm{m}}\right]^T$. The location of the target and the communication user are generated in simulation based on a uniform point distribution. The MSE of the MLE is evaluated using Monte Carlo simulations with 100 runs. The simulation related parameters are shown in Table \uppercase\expandafter{\romannumeral2}.
\begin{table}[t]
\centering
\caption{Simulation parameters}
\label{my-label}
\begin{tabular}{|c|c|c|c|}
\hline 
 \textbf{parameter}&\textbf{value}&\textbf{parameter}&\textbf{value}\\
 \hline 
 $\alpha_0$ & -50\,$\rm{dB}$ & $\beta_0$ & -47\,$\rm{dB}$  \\
 \hline 
 $N_0$& -170\,${\rm{dBm/Hz}}$ & $\sigma^2_0$ & $N_0B$   \\
 \hline 
  P &20\,dBm & $B$ & 1\,MHz   \\
 \hline 
 $G_{\rm{p}}$ & $0.1B$ &${V_{\max }}$& 30 m/s\\
 \hline 
 $H$&200\,m&$T_{\rm{h}}$& 1\,s  \\
 \hline 
 $V_{\rm{str}}$ & 20 m/s& $\mu$ & 5 \\
 \hline
  $L_x$ & 1500 m& $L_y$ & 1500 m \\
 \hline
 $T_{\rm{f}}$& 1.5\,s & &\\
 \hline
\end{tabular}
\end{table}

\begin{figure}[t!]
\centering
\subfigure[Sensing]{
\begin{minipage}[t]{0.7\linewidth}
\centering
\includegraphics[width=1\linewidth]{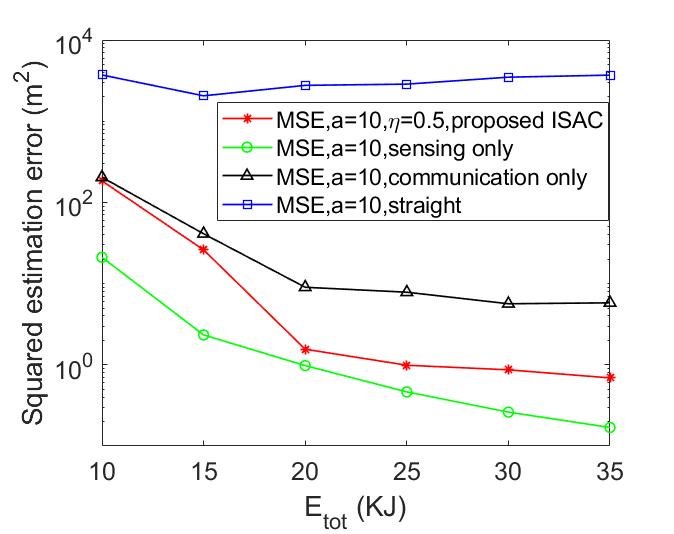}
\end{minipage}
}
\centering
\subfigure[Communication]{
\begin{minipage}[t]{0.7\linewidth}
\centering
\includegraphics[width=1\linewidth]{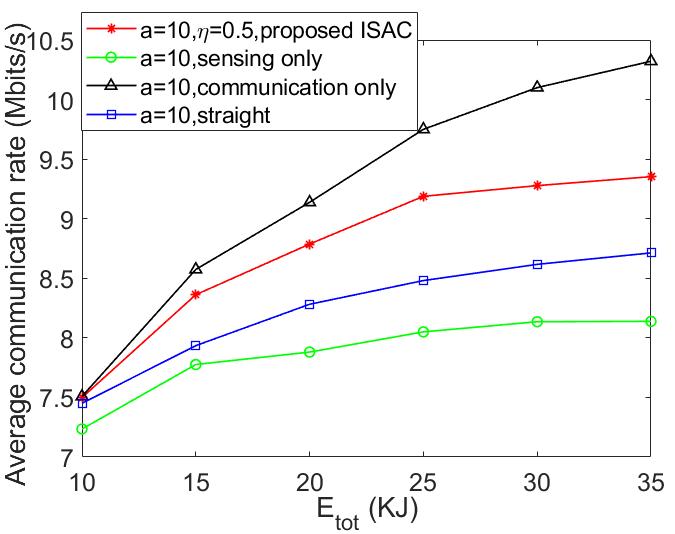}
\end{minipage}
}
\centering
\subfigure[Hovering points]{
\begin{minipage}[t]{0.7\linewidth}
\centering
\includegraphics[width=1\linewidth]{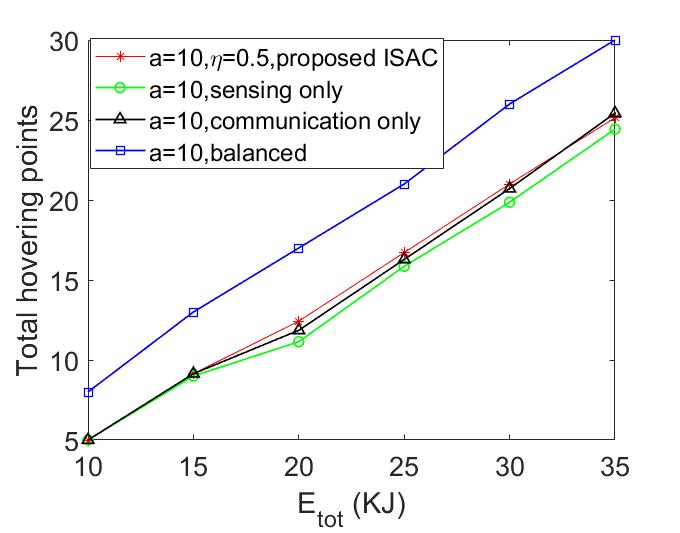}
\end{minipage}
}
\centering
\caption{Performances of different approaches versus $E_{{\rm{tot}}}$ ($N_{\rm{stg}}=25$).}
\end{figure}
\indent First, we study the convergence behaviour of the proposed iterative algorithm used in addressing ${\rm{P2'''}}\left(m\right)$. In Fig. 4, we show the convergence plot of the two performance metrics in the $1$st stage. We can observe that the iterative algorithm has a fast convergence rate reaching within 1\% of the final value in 5 iterations. In Fig. 5, we use ``sensing only'' to denote the trajectory design problem in ${\rm{P2}}\left(m\right)$ with $\eta=1$, so that we can evaluate the UAV sensing performance based on the proposed approaches with different measurement noise. We use ``MSE'' to represent the MSE of the target's coordinates estimate, and ``CRB'' is the CRB of the target's coordinates. It is shown that when $a=50$, which means when the measurement noise has a larger variance, compared with $a=10$, the estimation error increases. We also observe that as $E_{\rm{tot}}$ increases, both the ``MSE'' and  ``CRB'' decrease, as expected.

\subsection{The C$\&$S Performance of the Proposed ISAC Approaches and Benchmarks with UAV Energy Constraints}
\indent In Fig. 6, we compare the C\&S performance of the proposed approach with several benchmark schemes. ``Communication only'' denotes the trajectory design problem in focusing only on the communication performance, which is equivalent to set $\eta=0$ in the optimization problem ${\rm{P2}}\left(m\right)$. The UAV trajectory obtained by the proposed approach only consider communication user's location. The term ``straight'' represents a path that the UAV flies to the middle point between the target and communication user, which is illustrated as $\left[x^m_{\rm{mdl}},y^m_{\rm{mld}}\right]^T$ in Section \uppercase\expandafter{\romannumeral4}, and then, hovers there until the energy runs out. We first illustrate the C\&S performance with different energy constraints $E_{\rm{tot}}$. Fig. 6 (a) shows that the target estimation error obtained by the proposed approach is about an order of magnitude better than a communication-only approach, and more than two orders of magnitude of the ``straight''. Meanwhile, the trade-off design results in a performance loss in communication part comparing with communication-only approach. The growth of the total energy budget results in a performance improvement for sensing. Fig. 6 (b) shows that the average communication rate of the communication user increases with the increase of $E_{\rm{tot}}$, resulting in a better communication performance. \\
\begin{figure}[t!]
\centering
\subfigure[$E_{{\rm{tot}}}=15{\rm{KJ}}$]{
\begin{minipage}[t]{0.45\linewidth}
\centering
\includegraphics[width=1.1\linewidth]{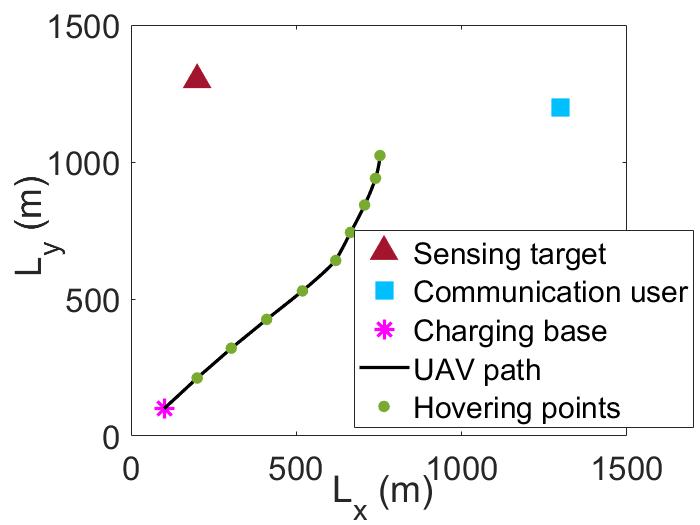}
\end{minipage}
}
\centering
\subfigure[$E_{{\rm{tot}}}=20{\rm{KJ}}$]{
\begin{minipage}[t]{0.45\linewidth}
\centering
\includegraphics[width=1.1\linewidth]{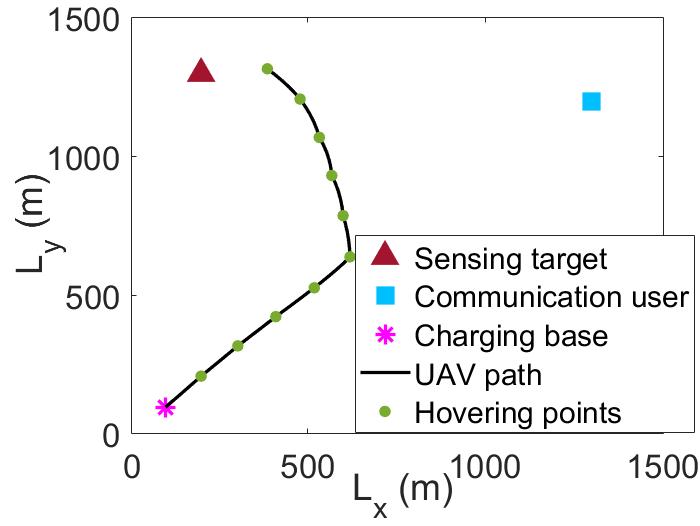}
\end{minipage}
}
\centering
\subfigure[$E_{{\rm{tot}}}=25{\rm{KJ}}$]{
\begin{minipage}[t]{0.45\linewidth}
\centering
\includegraphics[width=1.1\linewidth]{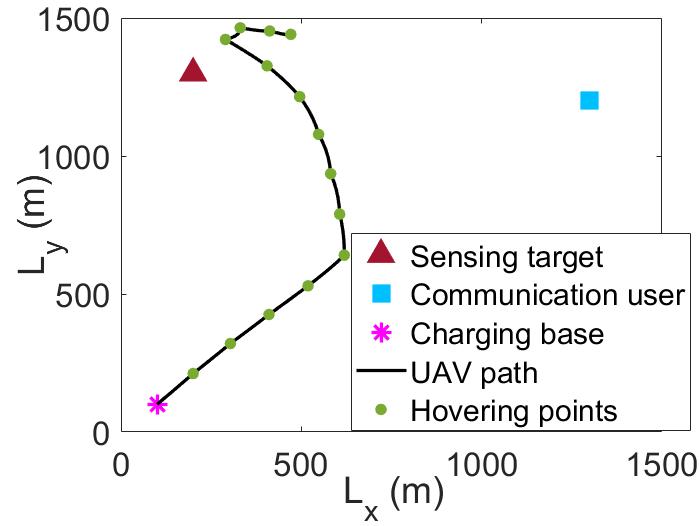}
\end{minipage}
}
\centering
\subfigure[$E_{{\rm{tot}}}=30{\rm{KJ}}$]{
\begin{minipage}[t]{0.45\linewidth}
\centering
\includegraphics[width=1.1\linewidth]{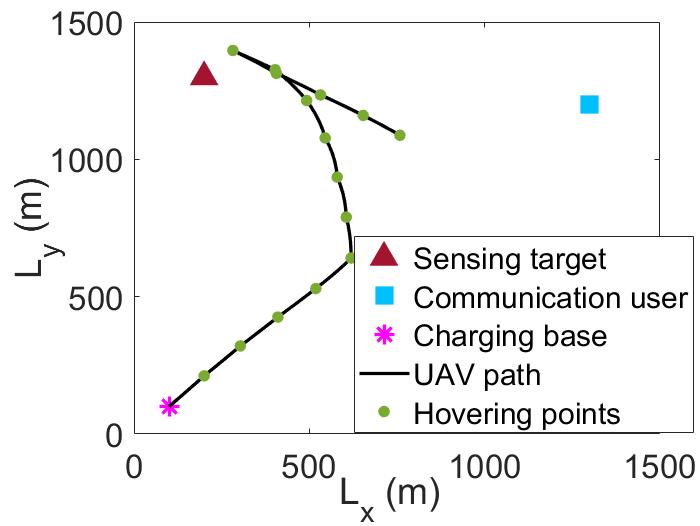}
\end{minipage}
}
\centering
\caption{UAV paths of the proposed trajectory design approach with different $E_{{\rm{tot}}}$ ($N_{\rm{stg}}=25$, $a=10$ and $\eta=0.5$).}
\end{figure}
\indent In Fig. 6 (c), larger $E_{\rm{tot}}$ means that the UAV hovers at more points and receives more measurements for sensing. Even thought there are more hovering points in ``straight'', if the UAV can not change its location to transmit to the user and sense the target, there is less contribution to the C\&S performance comparing with the proposed trajectory design.\\
\indent Fig. 7 compares the UAV paths with different $E_{\rm{tot}}$ to explain the reason that larger $E_{\rm{tot}}$ can achieve a better C$\&$S performance. Specifically, we observe that the UAV can fly a longer distance with larger $E_{\rm{tot}}$, which means more observation measurements can be obtained. This thus leads to a better sensing performance. We can also observe that with such a distribution of the target and the communication user, in the trade-off problem, sensing performance acquires a higher priority so that the UAV flies closer to the target. But, at the same time, the average distance between the UAV and the communication user decreases as $E_{\rm{tot}}$ increases, thus, a better communication performance is obtained. The practical trajectory is dependent on the distribution of the target and the communication user. For example, if the target and the communication user are close to each other, the UAV can fly closer to both the user and the target with large $E_{\rm{tot}}$.
\begin{figure}[t!]
\centering
\subfigure[$E_{\rm{tot}}=10{\rm{KJ}}$]{
\begin{minipage}[t]{0.7\linewidth}
\centering
\includegraphics[width=1\linewidth]{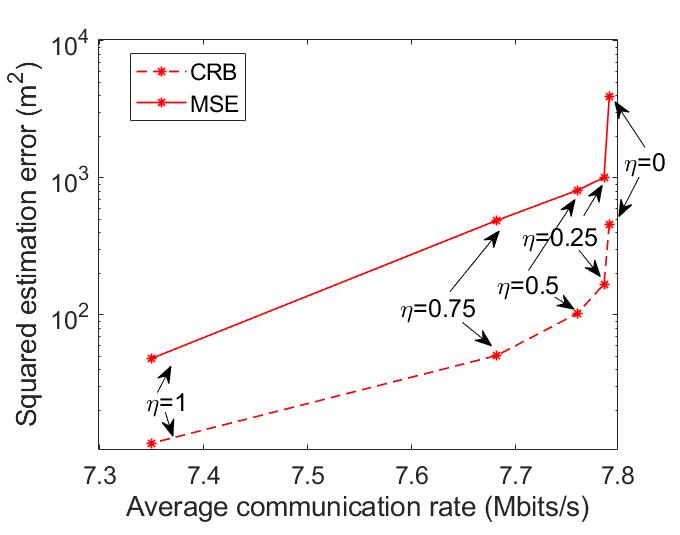}
\end{minipage}
}
\centering
\subfigure[$E_{{\rm{tot}}}=30{\rm{KJ}}$]{
\begin{minipage}[t]{0.7\linewidth}
\centering
\includegraphics[width=1\linewidth]{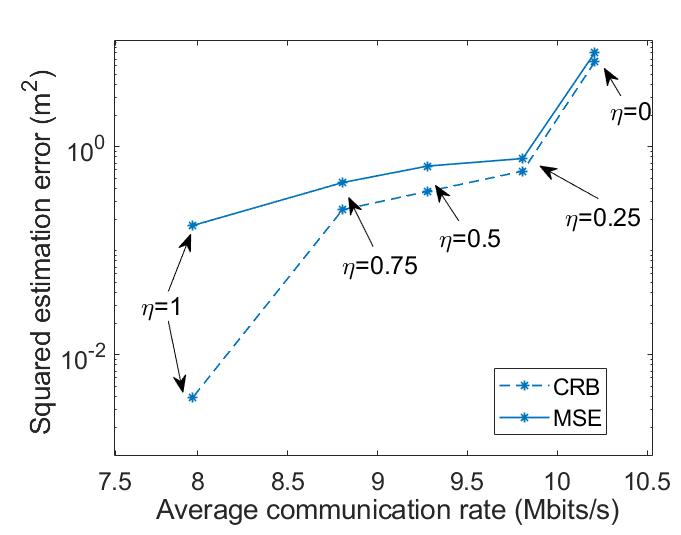}
\end{minipage}
}
\centering
\subfigure[UAV trajectories]{
\begin{minipage}[t]{0.8\linewidth}
\centering
\includegraphics[width=1\linewidth]{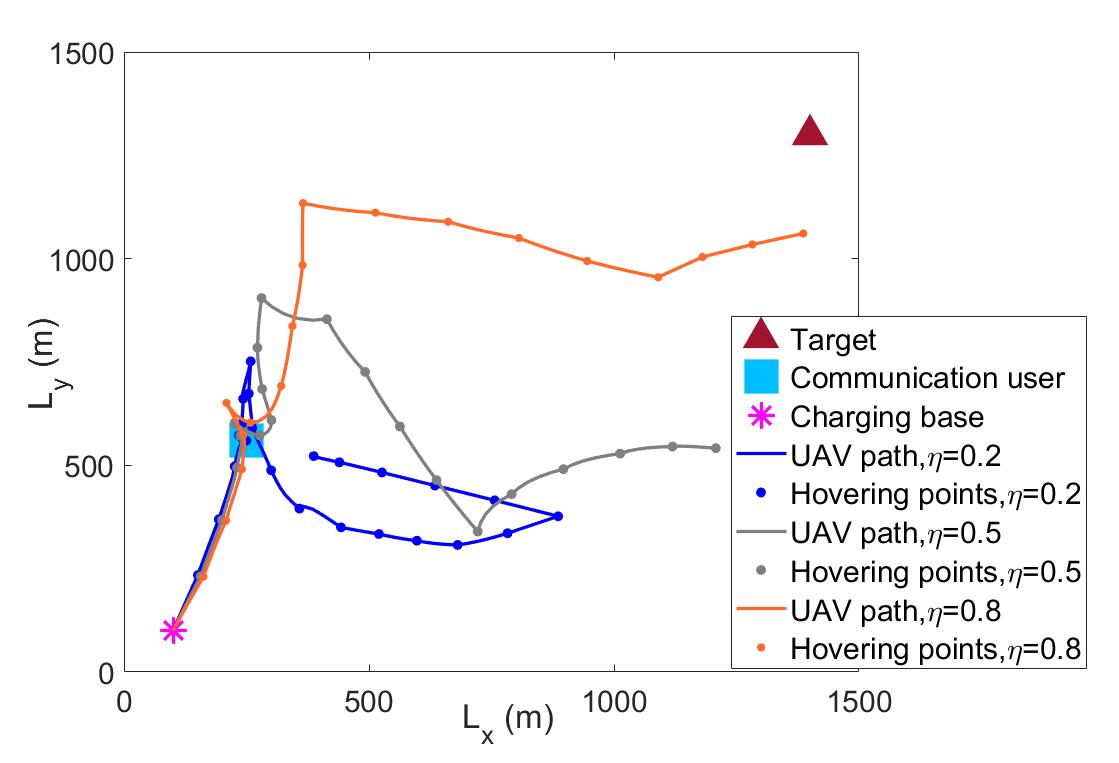}
\end{minipage}
}
\centering
\caption{Trade-off between C\&S in the proposed trajectory design approach ($N_{\rm{stg}}=25$).}
\end{figure}
\begin{figure}[t!]
\centering
\includegraphics[width=0.7\linewidth]{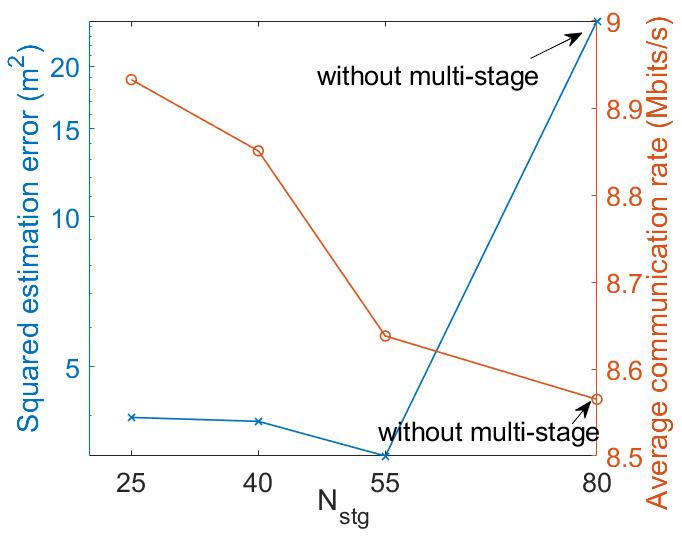}
\caption{Performance metrics of the proposed multi-stage approach with
different $N_{\rm{stg}}$ ($E_{\rm{tot}} = 40$KJ and $\eta=0.5$).}
\end{figure}
\subsection{The Trade-off between C\&S}

In Fig. 8, we aim to show the performance trade-off between C\&S , by tuning the weighting factor $\eta$. It can be seen that there exists a trade-off between the average communication rate and estimation error, especially when the $E_{\rm{tot}}$ is small. The C$\&$S performance changes smoothly from $\eta=0.25$ to $\eta=1$, but changes sharply from $\eta=0$ to $\eta=0.25$, which means that as long as the sensing is considered in the trajectory design problem, even if the weight is small, the sensing performance can result in a huge enhancement with a slight decrease on communication performance than ``communication only'' trajectory design.\\
\indent Fig. 8 (c) shows the UAV paths with different $\eta$ to explicate the trade-off between C\&S. It can be observed that with a larger $\eta$, since the sensing metric can contribute more to the objective value, the UAV trajectory is more suitable for performing sensing. For example, the UAV flies closer to the target, and the angle distribution between the target and the UAV hovering points is more diverse, so that the sensing can be performed from more directions, resulting in a better estimation performance.

\subsection{Behaviour of the Multi-stage Trajectory Design Approach}
In Fig. 9, we show how the performance of both C\&S is influenced by the proposed multi-stage approach. Here, we set the total number of the UAV waypoints is $N_1+N_2+...+N_M=N_{\rm{tot}}=80$, regardless of $E_{\rm{tot}}$, so that we can keep a fairness that both the proposed multi-stage approach and the trajectory design which performs with one time can have a same number of UAV waypoints and hovering points. If we set $N_{\rm{stg}}=25$, then, $M=4$. By applying the proposed multi-stage approach, in the stage 1, 2 and 3, $N_{\rm{stg}}=25$ waypoints are determined respectively. In the $4$-th stage, $N_{\rm{lst}}=5$ waypoints are determined. ``One-stage'' is the benchmark here, which means that all 80 UAV waypoints are designed at once ($N_{\rm{stg}}=80,M=1$) rather than being divided into several stages, so that there is no update for target's coordinates estimate. Fig. 9 shows that with multiple stages, the UAV trajectory design can lead to an enhancement for sensing performance. At the same time, the communication performance decreases in multi-stage approach. This is because that with an updated target location estimate, we can gradually obtain a more accurate target location, then the UAV trajectory design is more suitable for the sensing performance. With $N_{\rm{stg}}$ increases, the estimation error decreases, since when we have more observations in one stage, a better estimation result may be obtained. But we should notice that, in some cases, small $N_{\rm{stg}}$ means that we have more times to update a more accurate estimate. Therefore, there is a trade-off between more stages but less observations in each stage, and less stages but more observations in each stage, that depends on the exact case.
\begin{figure}[t]
\centering
\includegraphics[width=0.8\linewidth]{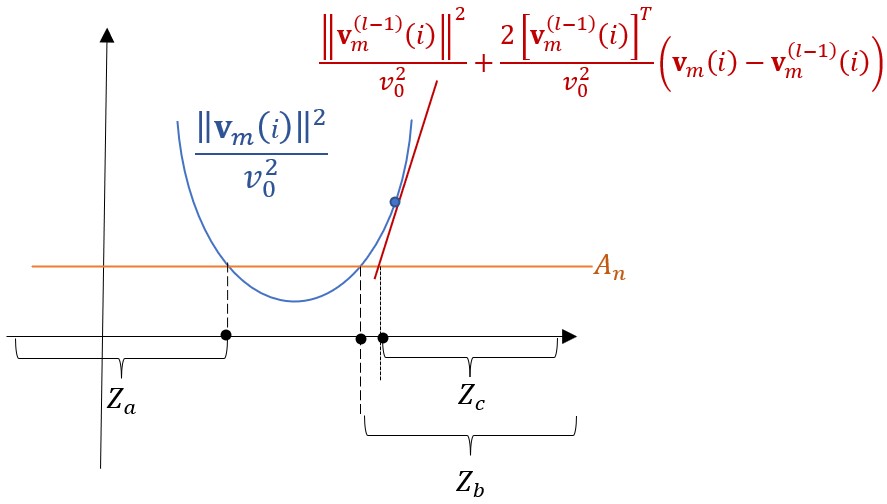}
\caption{Constraint (61).}
\end{figure}

\section{Conclusion}
In this paper, we have considered the UAV trajectory design and target estimation problem in a UAV based ISAC system. Firstly, we formulated a weighted optimization problem for the UAV trajectory design to achieve a flexible performance trade-off between C$\&$S. We further developed an iterative algorithm to solve the formulated problem. A multi-stage approach has been proposed to improve the sensing performance and obtain more accurate target's coordinates. The simulations have shown that the UAV energy capacity and the weighting factor have significant impact on the C$\&$S performance. Moreover, the trade-off designs can considerably improve both the C\&S performance comparing with a simple straight path. Future research will focus on multi-UAV trajectory design in ISAC system. Meanwhile, moving communication users and moving targets will be considered.

\begin{appendices}
\section{The proof about the feasibility of the solutions of $\left\{{\bf{S}}_m, {\bf{x}}_m,{\bf{y}}_m,{\bf{V}}_m,{\bm{\delta}}_m,{\bm{\xi}}_m\right\}$ after constraints apply SCA in ${\rm{P2}}\left(m\right)$}

\indent Since we seek the sub-optimal solution of $\left\{{\bf{S}}_m, {\bf{x}}_m,{\bf{y}}_m,{\bf{V}}_m,{\bm{\delta}}_m,{\bm{\xi}}_m\right\}$ following the descent direction in Section \uppercase\expandafter{\romannumeral4}, we need to prove that all available points belonging to the descent direction can satisfy all constraints in ${\rm{P2}}\left(m\right)$ after we use SCA for non-convex constraints. We rewrite ${\rm{P2}}\left(m\right)$ as
\begin{align}
    {\rm{P3}}\left(m\right):&\mathop {{\rm{min}}}\limits_{\left\{ {\bf{S}}_m, {\bf{x}}_m,{\bf{y}}_m,{\bf{V}}_m,{\bm{{\bf{\delta}}}}_m,{\bm{{\bf{\xi}}}}_m\right\}}\eta\cdot\widetilde{\rm{CRB}}^m _{x_{\rm{t}},y_{\rm{t}}} -\left(1-\eta\right)\cdot\overline{R}_m \nonumber\\
    &{\rm{s.t.}}\nonumber\\
    &\frac{\|{\bf{v}}_m\left(i\right)\|^2}{v_0^2}\geq\frac{1}{\left[\delta_m\left(i\right)\right]^2}-\xi_m\left(i\right), i=1,2,...,N_m,\nonumber\\
    &\left[\delta_m\left(i\right)\right]^2\geq \xi_m\left(i\right),i=1,2,...,N_m,\nonumber\\
    &\xi_m\left(i\right)\geq 0,i=1,2,...,N_m,\nonumber\\
    &{\rm{(40a),\ (40b),\ (42)\ and\ (44)}}.\nonumber
\end{align}
It is readily observed that all constraints are convex except of the first and the second constraints. Therefore, we need to prove that in the $l$-th iteration, all possible solution of $\left\{{\bf{V}}_m,{\bm{\delta}}_m,{\bm{\xi}}_m\right\}$ following the descent gradient still satisfies the first two constraints.\\
\indent First of all, we rewrite the first constraint into two constraints as follows
\begin{equation}
    \frac{\|{\bf{v}}_m\left(i\right)\|^2}{v_0^2}\geq A_m\left(i\right), i=1,2,...,N_m,
\end{equation}
\begin{equation}
    A_m\left(i\right)\geq \frac{1}{\left[\delta_m\left(i\right)\right]^2}-\xi_m\left(i\right), i=1,2,...,N_m.
\end{equation}
In Fig. 10, we assume that the set of $\|{\bf{v}}_m\left(i\right)\|$ satisfying (61) in the $l$-th iteration are sets ${\bm{Z}}_a$ and ${\bm{Z}}_b$. After applying the first-order Taylor expansion of (61), the available set for $\|{\bf{v}}_m\left(i\right)\|$ in the $l$-th iteration is ${\bm{Z}}_c$. Thus, the solution set ${\bm{Z}}_c$ now is included by ${\bm{Z}}_b$, which means, all possible solutions via SCA for $\|{\bf{v}}_m\left(i\right)\|$ in the $l$-th iteration belongs to its practical set and satisfy the first constraint in ${\rm{P3}}\left(m\right)$.  Therefore, all possible solutions of $\|{\bf{v}}_m\left(i\right)\|$ in the $l$-th iteration are in a feasible region. By a similar process, we can show that the feasibility regions are satisfied by all other optimization variables, omitted here for brevity.

\end{appendices}

\balance

\end{document}